\documentclass[12pt,english,american,peerreview]{IEEEtran}
\usepackage{cite}
\usepackage[T1]{fontenc}
\usepackage[latin9]{inputenc}
\usepackage[normalem]{ulem}

\usepackage{mathtools}
\usepackage{tabulary}
\usepackage{booktabs}
\usepackage{mathrsfs}
\usepackage{longtable}
\usepackage{mathalfa}
\usepackage{amsthm}
\usepackage{amsmath}
\usepackage{amssymb}
\usepackage{textcomp}
\usepackage{graphicx}
\usepackage{slashbox}
\usepackage{mathrsfs}

\usepackage{enumitem}
\usepackage{relsize}
\usepackage{caption}
\makeatletter

\newcommand{\suchthat}{\;\ifnum\currentgrouptype=16 \middle\fi|\;}
\newcommand{\mc}{\mathcal}
\newcommand{\mb}{\mathbf}

\newcommand{\mbb}{\mathbb}

\newcommand{\pmap}{$p$-map}

\newcommand{\spc}{\hspace{1mm}}

\newcommand{\tsf}{\textsf}
\newcommand{\bs}{\boldsymbol}

\theoremstyle{plain}

\theoremstyle{plain}
\newtheorem{defn}{\protect\definitionname}
\theoremstyle{plain}
\newtheorem{lem}{\protect\lemmaname}
\theoremstyle{plain}

\theoremstyle{plain}

\theoremstyle{definition}
\newtheorem{exmp}{Example}

\@ifundefined{definecolor}{\usepackage{color}}{}

\usepackage{babel}\providecommand{\definitionname}{Definition}

\makeatother

\usepackage{babel}
\addto\captionsamerican{\renewcommand{\propname}{Proposition}}
\addto\captionsamerican{\renewcommand{\corollaryname}{Corollary}}
\addto\captionsamerican{\renewcommand{\definitionname}{Definition}}
\addto\captionsamerican{\renewcommand{\lemmaname}{Lemma}}
\addto\captionsamerican{\renewcommand{\theoremname}{Theorem}}
\addto\captionsenglish{\renewcommand{\propname}{Proposition}}
\addto\captionsenglish{\renewcommand{\corollaryname}{Corollary}}
\addto\captionsenglish{\renewcommand{\definitionname}{Definition}}
\addto\captionsenglish{\renewcommand{\lemmaname}{Lemma}}
\addto\captionsenglish{\renewcommand{\theoremname}{Theorem}}
\providecommand{\corollaryname}{Corollary}
\providecommand{\definitionname}{Definition}
\providecommand{\lemmaname}{Lemma}
\providecommand{\theoremname}{Theorem}
\providecommand{\propname}{Proposition}
\usepackage[lined,linesnumbered]{algorithm2e}
\makeatletter
\@ifpackageloaded{hyperref}{%
  \def\elabel#1{\label{#1}}%
}{%
  \def\elabel#1{\@bsphack \protected@write \@auxout {}{\string \newlabel {#1}{{\theAlgoLine }{\thepage }}}\@esphack}%
}
\makeatother
\usepackage[normal]{subfig}

\usepackage{color}
\definecolor{FireBrick}{rgb}{0.5812,0.0074,0.0083}
\definecolor{RoyalBlue}{rgb}{0.0236,0.0894,0.6179}
\definecolor{RoyalGreen}{rgb}{0.0236,0.6179,0.0894}
\definecolor{RoyalRed}{rgb}{0.6179,0.0236,0.0894}
\definecolor{LightBlue}{rgb}{0.8544,0.9511,1.0000}
\definecolor{Black}{rgb}{0.0,0.0,0.0}

\definecolor{linkColor}{rgb}{0.0,0.0,0.554}
\definecolor{citeColor}{rgb}{0.0,0.0,0.554}
\definecolor{fileColor}{rgb}{0.0,0.0,0.554}
\definecolor{urlColor}{rgb}{0.0,0.0,0.554}
\definecolor{promptColor}{rgb}{0.0,0.0,0.589}
\definecolor{brkpromptColor}{rgb}{0.589,0.0,0.0}
\definecolor{gapinputColor}{rgb}{0.589,0.0,0.0}
\definecolor{gapoutputColor}{rgb}{0.0,0.0,0.0}

\definecolor{FuncColor}{rgb}{0.0,0.0,0.0}
\definecolor{Chapter }{rgb}{0.0,0.0,0.0}
\definecolor{DarkOlive}{rgb}{0.1047,0.2412,0.0064}
\usepackage{fancyvrb}

\begin{document}

\title{Constrained Linear Representability of Polymatroids and Algorithms for Computing Achievability Proofs in Network Coding}
\author{Jayant Apte, \IEEEmembership{Member, IEEE,} John MacLaren Walsh, \IEEEmembership{Member, IEEE}
\thanks{Support under National Science Foundation awards CCF-1016588 and CCF-1421828 is gratefully acknowledged. Jayant Apte and John MacLaren Walsh are with the Department of Electrical and Computer Engineering, Drexel University, Philadelphia, PA, USA (email: jsa46@drexel.edu, and jwalsh@coe.drexel.edu). Preliminary ideas related to this work were presented at ISIT 2014\cite{Apteisit14}.}}

\maketitle
\noindent\begin{abstract}
The constrained linear representability problem (CLRP) for polymatroids determines whether there exists a polymatroid that is linear over a specified field while satisfying a collection of constraints on the rank function. Using a computer to test whether a certain rate vector is achievable with vector linear network codes for a multi-source network coding instance and whether there exists a multi-linear secret sharing scheme achieving a specified information ratio for a given secret sharing instance are shown to be special
cases of CLRP. Methods for solving CLRP built from group theoretic techniques for combinatorial generation are developed and described.  These techniques form the core of an information theoretic achievability prover, an implementation accompanies the article, and several computational experiments with interesting instances of network coding and secret sharing demonstrating the utility of the method are provided.
\noindent \keywords network coding, secret sharing, polymatroids, entropy function, computer assisted achievability proofs
\end{abstract}

\section{Introduction}
For many important multiterminal information theory problems, including those arising in network coding, distributed storage, and secret sharing, using a computer to perform an arbitrary achievability proof requires one to know an algorithm to determine if there exists an almost entropic polymatroid satisfying certain constraints on its rank function.  Finding such an algorithm is a fundamental open problem in information theory, also known as the problem of characterization of the closure of the region of entropic vectors ($\overline{\Gamma_N^*}$).  We consider a special case of this very difficult problem, which we call the Constrained Linear Representability Problem (CLRP) for polymatroids.  We show that the ability to solve CLRP can automate the achievability proofs that one encounters while determining the performance of linear codes in multi-source multi-sink network coding over directed acyclic hypergraphs and in secret sharing and while proving new linear rank inequalities.

Traditionally, the achievability constructions in proofs for these problems are performed manually, which makes them tedious and time consuming. A computer program to solve CLRP, at least for small and moderate instances, enables one, in turn, to pursue a computational agenda for approaching problems like network coding and secret sharing, which have proven difficult to solve in general. This involves solving small instances of these problems to build large and exhaustive databases of solved instances which can then be analyzed to find patterns and structure that allow one to make much more general statements about these problems \cite{Li_PhDdissertation,Congduan_MDCS,Li_Operators}. All of these tasks would be impossible without the use of a computer due to sheer number of man hours required. 


This article develops and describes a method for solving constrained linear representability problems built from group theoretic techniques for combinatorial generation.  More precisely, in section \ref{sec:prelim}, after reviewing the definition of the region of entropic vectors and its linear polymatroid inner bound, we define two variants of CLRP, one existential and one enumerative.  Then, Section \ref{sec:clrpncss} shows in detail how the problems of calculating achievability proofs for fundamental limits for network coding rate regions and secret sharing can be viewed as instances of these CLRPs.  Section \ref{sec:ingredients} then defines key concepts and terminology which enable techniques developed for combinatorial generation to be applied to CLRP, explaining along the way key decisions enabling CLRP to be solved in an efficient manner that can exploit problem symmetry and handle isomorphism.  Building upon these ideas, section \ref{sec:polyext}
presents the developed algorithm for solving CLRP, which is also implemented as a GAP package the \tsf{I}nformation \tsf{T}heoretic \tsf{A}chievability \tsf{P}rover-- \tsf{ITAP}, the first of its kind, accompanying the article.  Finally, Section \ref{sec:example}
describes several quantities playing a key role in the complexity of the developed method, then provides a series of examples of achievability problems solved with ITAP.

\section{Entropy, Polymatroids, and Constrained Linear Representability Problems}\label{sec:prelim}
In this section, we introduce basic terminology and review several key existential problems related to the region of entropic vectors in \S\ref{sub2a}.  We also state the existential and enumerative variants of the constrained linear representability problem (CLRP), which is the main subject of this paper.  The relevance of the two variants of CLRP stated in this section to network coding and secret sharing will be discussed in \S\ref{sec:clrpncss}, along with a review of prior work for solving these existential questions. For the basic terminology related to matroid theory and coding theory, we refer to the books by Oxley \cite{OxleyMatroidBook} and Betten et al. \cite{betten2006error} respectively.  The commonly used symbols and notation are described in table \ref{tab:TableOfNotationForMyResearch}. 

Consider a collection of $N$ discrete random variables $\boldsymbol{X}_N=(X_1,\hdots,X_N)$. The associated \textit{entropy vector}
is a $2^N-1$-dimensional vector obtained by stacking the entropies of subsets $\boldsymbol{X}_A\triangleq (X_n|n\in A)$ of $\boldsymbol{X}_N$ into a
vector $\mb h\triangleq (h(\boldsymbol{X}_{A})\suchthat  A\subseteq [N])$. By convention, $h(\emptyset)=0$.  Let $\mc D_N$ be the set of all
joint probability mass functions for $N$ discrete random variables. Then the \textit{entropy function} is the map $\mb h :\mc D_N\rightarrow \mbb R^{2^N-1}$ mapping a joint probability mass function to its entropy vector.  Now consider the following problem:
\begin{itemize}
  \item{(E1)} Given a vector $\mb h'\in \mbb R^{2^N-1}$, determine whether $\mb h'=\mb h(p)$ for some $ p\in \mc D_N$.
\end{itemize}
If the answer to E1 is 'yes', then there exists a joint probability mass function $p$ s.t. $\mb h'=\mb h(p)$, and $\mb h'$ is said to be \textit{entropic}. We can now define the region of entropic vectors $\Gamma_N^*$ as
\begin{equation}
\Gamma_N^*\triangleq \{\mb h\in \mbb R^{2^N-1}\suchthat \mb h \text{ is entropic }\}
\end{equation}
Based on above definition, problem (E1) can be seen to be equivalent to testing membership in $\Gamma_N^*$. The closure of $\Gamma_N^*$ is a convex cone. Characterization of $\Gamma_N^*$ and $\overline{\Gamma_N^*}$ is of central importance in network information theory as they determine all fundamental information inequalities \cite{YeungBook}, limits for secret sharing systems \cite{beimelsurvey11}, and the capacity regions of all networks under network coding \cite{yanmultisourceTIT,Li_Operators}.

\begin{longtable}{| p{.20\textwidth} | p{.80\textwidth} |}  
\caption{Notation}\label{tab:TableOfNotationForMyResearch}\\

\hline
$ \setminus$ & Set difference or polymatroid deletion\\
$\leq$ & Subgroup relationship or inequality\\
$\boldsymbol{X}_N$ & Set of $N$ discrete RVs\\
$[N]$ & The set $1,\hdots, N$ for $N\in\mbb N$\\
\vtop{\hbox{$\mathscr{A,B,C}$}\hbox{$\mathscr{D,S,I}$}} & Generic sets or multisets\\
$E$& Ground set of a polymatroid (usually same as $[N]$)\\
$\bs X_A$& Subset of $\bs X_N$\\
$\mb h$& A vector in $\mbb R^{2^N-1}$ or the entropy function, depending on context\\
$\mb h_{\mathscr S}$& Entry in vector $\mb h$ corresponding to subset $\mathscr S$ or entropy of a subset $\mathscr S$ of $\bs X_N$\\
$\Gamma_N^*$& Set of entropic vectors\\
$\overline{\Gamma_N^*}$& Closure of set of entropic vectors\\
$\mc D_N$& Set of all joint distributions on $\bs X_N$\\
$\tsf{rk},f$ & set functions\\
$2^{\mathscr S}$ & Power set of the set $\mathscr S$\\
$c$ & A constant\\
$P$ & A polymatroid, i.e. $2$-tuple consisting a set and a set function\\
$P^c$ & Polymatroid $P$ with rank function scaled by $c$\\
$\mbb K$ & An arbitrary field\\
$q$ & A prime power\\
$\mbb F_q$ & The $q$-element finite field\\
$\Gamma_N^{\tsf{space}}$ & Subspace inner bound\\
$\beta$ & demands of sink nodes\\
$e,\mathscr E$ &  a (hyper)edge or an encoder, set of all (hyper)edges or encoders\\
$\mathscr F$ &  head nodes of a hyperedge\\
$g,\mathscr G$ &  an intermediate node, set of intermediate nodes\\
$s,\mathscr S$ &  a source node, set of source nodes\\
$t,\mathscr T$ &  a sink node, set of sink nodes\\
$\mc I$ &  A collection of constraints on the polymatroid rank function\\
$\mathfrak P$ &  A list of  polymatroids\\
$\tsf{rep}(i)$ &  Bases of vector space associated with ground set element $i$\\
$p$-map &  An partial map from ground set of a polymatroid to $\bs X_N$\\
$\phi$ &  An isomorphism or a $p$-map\\
$V_i,W_i$ &  Subspace associated with ground set element $i$, presented as a matrix with $\tsf{rep}(i)$ as columns\\
$\equiv$ & An arbitrary equivalence relation on a set of polymatroids or their representations\\
$\cong$ &  Strong isomorphism relation between polymatroids or their representations\\
$\overset{W}{=}$ &  Weak isomorphism relation between polymatroids or their representations\\
$U^i_j$ &  Uniform matroid of rank $i$ and ground set size $j$\\
$\mb u,\mb v$ & arbitrary vectors\\
$\tsf{Gr}_q(r,k)$ &  Set of all $k$ dimensional subspaces of $\mbb F_q^r$\\
$\tsf{Gr}_q(r,K)$ &  Union of several $\tsf{Gr}_q(r,k),k\in K$\\
$GL(r,q)$ &  General linear group associated with $\mbb F_q^r$\\
$\tsf{Gal(q)}$ &  Galois group of $\mbb F_q$ which is the group of all automorphisms of $\mbb F_q$\\
$\Gamma L(r,q)$ &  General semilinear group associated with $\mbb F_q^r$\\
$PGL(r,q)$ &  Projective general linear group associated with $\mbb F_q^r$\\
$P\Gamma L(r,q)$ &  Projective semilinear group associated with $\mbb F_q^r$\\
$\mc Z_r$ &  Group formed by scaled versions of $\mbb I_r$ over $\mbb F_q$\\
$\langle \mb v_1,\hdots,\mb v_N\rangle$ &  Vector subspace generated by vectors $\mb v_1,\hdots,\mb v_N$\\
$K_P$ & Set of distinct singleton ranks associated with polymatroid $P$\\
$\mathscr S_{N,r,K}$ & Set of all representations of simple polymatroids of rank $r$, size $N$ with singleton ranks from set $K$\\
$\mathscr P^q(\mb c)$ & A class of linear codes defined by the class tuple $\mb c$\\
$\mathscr P^q(\mb c,A)$ & Linear network codes for HMSNC instance $A$ from class $\mathscr P^q(\mb c)$\\
$\mc R^*$ & The exact rate region of a HMSNC instance\\
$\mc R_{\tsf{out}}$ & An outer bound on the exact rate region of a HMSNC instance\\
$\mc R_{\tsf{in}}$ & An inner bound on the exact rate region of a HMSNC instance\\
\bottomrule

\end{longtable}

Fujishige \cite{Fujishige_IC_78} observed that a vector $\mb h\in\Gamma_N^*$ must satisfy the polymatroidal axioms (P1)-(P3) in the definition below.
\begin{defn}
  A polymatroid $(E,f)$ consists of a set $E$, called the ground set, and a set function $f:2^E\rightarrow \mbb R$, called the rank function, which satisfies the following properties:
  \begin{itemize}
    \item{[(P1)]} $f(\emptyset)=0$ (normalized)
    \item{[(P2)]} $f(\mathscr A)\geq f(\mathscr B),\forall  \mathscr B\subseteq \mathscr A\subseteq E$ (monotone)
    \item{[(P3)]} $f(\mathscr C)+f(\mathscr D)\geq f(\mathscr C\cup \mathscr D) + f(\mathscr C \cap \mathscr D), \ \forall \mathscr C,\mathscr D \subseteq E$ (submodular)
    \end{itemize}
\end{defn}
Note that a polymatroid $(E,f)$ is a matroid if it is integer valued, $f:2^E \rightarrow \mbb Z_{\geq 0}$, and $f(i)\leq 1$ for each $i\in E$. (P1)-(P3) generate all Shannon type inequalities\cite{YeungBook}. Unfortunately, Shannon type inequalities are only necessary conditions
for determining whether or not a candidate vector is entropic. Inequalities not implied by (P1)-(P3) that are satisfied by all entropic vectors exist and are called non-Shannon type inequalities.
The first such inequality was found by Zhang and Yeung \cite{Zhang_TIT_07_98}.  Hundreds of linear non-Shannon type inequalities have been
found since the Zhang-Yeung inequality \cite{dfz6new,Csirmaz2013InforIneqfor4}. Furthermore, Mat\'us \cite{Matus_ISIT_2007} has shown that for $N\geq 4$, an infinite number of linear information inequalities is necessary to determine $\bar{\Gamma}^*_N$.

\subsection{Constrained Linear Representability Problems for polymatroids}\label{sub2a}
The \textit{representable} polymatroids are a sub-class of entropic polymatroids that arise from subspace arrangements. 

\begin{defn} \label{reppolydef}
A subspace arrangement in a $k$-dimensional vector space $V=\mbb K^k$ over some field $\mbb K$ is a multiset of $N$ subspaces $\mathscr S=\{V_1,V_2,\hdots,V_N\}$ with $V_n \subseteq V\ \forall n \in [N]$.
All vector spaces considered in this work are assumed to be finite dimensional.
The rank function of the subspace arrangement is $\tsf{rk}:2^{\mc [N]}\rightarrow \mbb Z$ defined as
\begin{equation}\label{eq:rkdef}
  \tsf{rk}(\mathscr I)\triangleq \tsf{dim}\sum_{i\in \mathscr I} V_i, \quad \forall \mathscr I\subseteq [N]
  \end{equation}
\end{defn}
Note that in \eqref{eq:rkdef} the sum is understood to be the direct sum of subspaces of a vector space.  As $([N],\tsf{rk})$ satisfies (P1)-(P3), it is a valid polymatroid. However, the converse does not necessarily hold true. This motivates the notion of representable polymatroids.
\begin{defn}\label{def:polyrep}
  A polymatroid $P=(E,f)$ is said to be representable if there exists a subspace arrangement $\mathscr S=\{V_1,\hdots,V_{\vert E\vert}\}$ over some field $\mbb K$
and a bijection $m:E\rightarrow [\vert E\vert]$ s.t. $f(\mathscr S)=\tsf{rk}(m(\mathscr S)),\spc\forall \mathscr S\subseteq E$.
\end{defn}
A theorem of Rado\cite{radorep} which states that a polymatroid is
representable over an infinite field then it is also representable over some finite field, allows us to restrict attention to fields $\mbb K$ with
only a finite number of elements when considering representability of polymatroids. In the above definition, if $\mbb K$ is a finite field with $q$ elements (i.e. $q$ is a prime power), then we say that the polymatroid is $\mbb F_q$-\textit{representable}. Note that it is possible for $P=(E,f)$ to not be $\mbb F_q$-representable
while a scaled version of $P$ i.e. $P^c=(E,cf), c>0$ is. We emphasize this distinction, so as to avoid confusion later.  One fact that makes representable polymatroids interesting is stated below. 
\begin{lem}(\hspace{-0.5mm}\cite{Hammer2000ShannEntr}, Thm. 2) \label{rep_ent}
 Every representable polymatroid is proportional to an entropic polymatroid.
\end{lem}
To see why the above statement is true, and to fix some notation, consider a representable polymatroid $P=([N],f)$ and let $\{V_1,\hdots,V_N\}$ be the associated subspace arrangement over finite field
$\mbb F_q$ with $q$ elements. Let the bijection mentioned in def. \ref{def:polyrep} be the identity map from $[N]$ to itself. We assume that each subspace $V_i,i\in [N]$ is presented as a set $\tsf{rep}(i)$ of $f(i)$ vectors in $\mbb F_q^{f([N])}$ forming a basis of the subspace $V_i$. Let $\mbb M(\mathscr I)$ be the matrix $[\tsf{rep}(i) | i\in \mathscr I]$  for each $\mathscr I\subseteq [N]$ (i.e. collect together the columns in $\tsf{rep}(i)$ for each $i\in \mathscr I$).  Consider a random row vector $\mb u \sim\mc U(\mbb F_q^{f([N])})$ uniformly distributed over $\mbb F_q^{f([N])}$, and create a collection of random variables
$\bs X_N=\{X_1,\hdots,X_N\}$ such that $X_i=\mb u\mbb M(i)$ is a random variable taking values in $\mbb F_q^{f(i)}$.  The entropy function maps $\bs X_N$ to a vector $\mb h$ such that for each $\mathscr A\subseteq [N]$, the entropy $\mb h_{\mathscr A}= f(\mathscr A)\log_2q=\tsf{rk}(\mbb M(\mathscr A))\log_2q$. The polymatroid $P^{\log_2 q}=([N],(\log_2 q) f)$ is then entropic by construction, thus completing the proof.

We now state the representability problem for integer polymatroids.
\begin{itemize}
  \item{[E2]} Given an integer polymatroid $P=(E,f)$, determine if there exists a representation of $P^{\alpha}$ over a finite field $\mbb F_q$ for some $\alpha\in \mbb N$.
\end{itemize}
Lemma \ref{rep_ent}, along with the fact that $\overline{\Gamma_N^*}$ is a convex cone allow us to construct an inner bound $\Gamma_N^{\tsf{space}}$ on $\overline{\Gamma_N^*}$, that we call the \textit{subspace inner bound}, which is the conic hull of all representable polymatroids. Problem (E2) is equivalent to testing membership of an integer polymatroid within an inner bound to $\Gamma_N^{\tsf{space}}$ formed by taking the conic hull of all polymatroids representable over a particular finite field.
 $\Gamma_N^{\tsf{space}}$ is polyhedral for $N\leq 5$\cite{DFZ2009Ineqfor5var}, while for $N\geq 6$, it remains
unknown whether this set is polyhedral. The linear inequalities that are true for all points in $\Gamma_N^{\tsf{space}}$ are called \textit{rank inequalities}.
For $N\leq 3$, the minimal (conic independent) set of rank inequalities is same as the minimal set of Shannon-type inequalities, while for $N=4$, the minimal set of rank inequalities is the minimal set of Shannon-type inequalities and all $6$ permutations of Ingleton's inequality\cite{Ingleton1971,Hammer2000ShannEntr}. For $N=5$, there are $24$ new linear rank inequalities that, together with the Shannon-type and Ingleton inequalities, form the minimal set of inequalities \cite{DFZ2009Ineqfor5var}.  One way to approach [E2] computationally, is to solve a restriction of [E2] for specific $\alpha$ and size of finite field $q$, iteratively for increasing values of $\alpha$ and $q$ until we find a representation. 
This leads us to the following problem:
\begin{itemize}
  \item{[E2$_q$]} Given a polymatroid $P=(E,f)$, determine if there exists a representation of $P$ over the finite field $\mathbb{F}_q$ with $q$ elements.
\end{itemize}
A special case of [E2$_q$] that is well-studied is the situation where $P$ is restricted to be a matroid.  A famous conjecture of Rota\cite{rotacon}, recently declared proven \cite{rotaproof}, states that for every finite field $\mbb F_q$, there are only a finite number of forbidden minors, a series of smaller matroids obtained through contraction and deletion, for a matroid to be representable over that field. 

We now define the notion of representation of a representable polymatroid.
\begin{defn} \label{def:reppoly}
 A representation of a $\mbb F_q$-representable polymatroid $([N],f)$ associated with a multiset $\mathscr S=\{V_1,\hdots,V_n\}$ of subspaces is a multiset of matrices $\{\mbb M(i),i\in[N]\}$ where each subspace $V_i$ is represented as a $f([N])\times f(i)$ matrix $\mbb M(i)$ whose columns are from the set $\tsf{rep}(i)$ for all $i\in [N]$.
 \end{defn}
In the context of the definition above, two representations of the same polymatroid are considered distinct if they correspond to different multisets of subspaces.  We will consider two other notions of distinctness via isomorphism later in the manuscript. Note that each subspace in such a multiset may be represented by several different bases. For the sake of simplicity, we do not distinguish between two representations that differ only in this sense. For the special case of matroids of ground set size $N$, we will assume that a representation is a multiset of vectors $\{\mb v_1,\hdots,\mb v_N\}$, each of which generates a subspace of dimension at most $1$.  Let $\tsf{Gr}_q(r,k),0\leq k\leq N$ be the set of all $k$-dimensional subspaces of $\mbb F_q^r$. The set $\tsf{Gr}_q(r,k)$ is also known as the Grassmannian. The size of $\tsf{Gr}_q(r,k)$ is given by the $q$-ary Gaussian binomial coefficient:
\begin{equation}
\vert \tsf{Gr}_q(r,k)\vert= {r\choose k}_q\triangleq \frac{(q^r-1)(q^{r-1}-1)\hdots(q^{r-k+1}-1)}{(q^k-1)(q^{k-1}-1)\hdots (q-1)}
\end{equation}
For a set $K\subseteq [r]$, define $\tsf{Gr}_q(r,K)\triangleq \bigcup_{k\in K} \tsf{Gr}_q(r,k)$.  We can associate with a polymatroid $P=([N],f)$, a set $K_P$ of distinct singleton ranks i.e. $K_P\triangleq \{f(i)\suchthat i\in [N]\}$.  In the spirit of compactness, it is also important to consider \emph{simple} polymatroids.  A matroid is said to be simple if it has no parallel elements or loops\cite{OxleyMatroidBook}, and likewise we have the following definition for a simple polymatroid
   \begin{defn}\label{nonsimplepoly}
     A polymatroid $(E,f)$ is said to be simple if it satisfies following conditions:
     \begin{enumerate}
       \item $\not\exists e_1,e_2\in E$ s.t. $f(\{e_1,e_2\})=f(e_1)=f(e_2)$
       \item For every $e\in E$, $f(e)>0$
     \end{enumerate}
   \end{defn}
   If there exist elements $e_1,e_2$ satisfying condition 1 of definition \ref{nonsimplepoly} above,
   they are said to be \textit{parallel} with respect to each other. In fact, there can exist multiple such elements forming a parallel class.
   \begin{defn}
     For a polymatroid $(E,f)$, $S\subseteq E$ is said to form a parallel class if
     \begin{equation}
       f(S)=f(s) \spc\forall s\in S
       \end{equation}
     In this case, $\vert S\vert$ is said to be the degree of the parallel class.
   \end{defn}
   If there exists an element violating condition 2 of definition \ref{nonsimplepoly}, then it is
   called a \textit{loop}. The number of ground set elements that are loops is called the loop degree
   of the polymatroid.  Given a polymatroid $P=(E',f')$ with an order on the ground set $E'$, we can associate it with a simple polymatroid
   $(E,f)$, s.t. $E\subseteq E'$ and $f(S)=f'(S), \forall S\subseteq E$ where $E$ is obtained by deleting all loops
   and all but one element of each parallel class from $E'$. Furthermore, we can decide to keep the smallest member of each parallel
   class under the aformentioned order on $E'$. We call $(E,f)$ obtained in this manner the \textit{unique simple polymatroid} associated with $(E',f')$, denoted as $\tsf{us}(P)$, and define, via an abuse of notation, $\vert \tsf{us}(P) \vert = \vert E \vert$. 
Having defined the parameters $K_P$ and $\tsf{us}(P)$ we can define the set $\mathscr P^q(N,r,K,s)$ as the set of representations of polymatroids with ground set size $N$, containing subspaces of $\mbb F_q^r$ s.t. $K_P\subseteq K$ and $\vert \tsf{us}(P)\vert =s$   over $\mbb F_q$. Finally, we define the set $\mathscr P^q(N,(r_l,r_u),K,(s_l,s_u) )$ of polymatroid representations as
\begin{equation}
\mathscr P^q(N,(r_l,r_u),K,(s_l,s_u) ) \triangleq \bigcup_{\begin{aligned}r_l \leq r\leq r_u\\ s_l\leq s\leq s_u\end{aligned} } \mathscr P^q(N,r,K,s)
\end{equation}
Henceforth, we shall refer to the tuple $\mb c=(N,(r_l,r_u),K,(s_l,s_u) )$ as the \textit{class tuple} and abbreviate $\mathscr P^q(N,(r_l,r_u),K,(s_l,s_u) )$  to $\mathscr P^q(\mb c)$.  This leads us to state precisely what we mean by a class of linear codes for the purpose of this work.
\begin{defn}
A class of linear codes  is any set $\mathscr P^q(\mb c)$ where $\mb c=(N,(r_l,r_u),K,(s_l,s_u) )$ is the class tuple satisfying  $r_l\leq r_u\leq N\cdot \max{K},s_l\leq s_u\leq N$ where $K$ is a finite set containing non-negative integers.
\end{defn} 
We are now ready to define the central problem adressed in this work. In several problems of practical interest, which will be described in the next section, rather than testing representability for a polymatroid specified by its rank function, it is of interest to determine whether there exists a polymatroid  representable over $\mbb F_q$ satisfying a specified collection of linear constraints on its rank function, and belonging to a particular class of linear codes.  We will name this problem the constrained linear representability problem (CLRP) for polymatroids.   More formally, let $\mc I$ be a collection of linear constraints on the rank function of a polymatroid with ground set $[N]$ and $\mb c$ be any valid class tuple. Then, the existential variant of CLRP can be stated as follows:


\begin{itemize}
  \item{[CLRP$_q$-EX]} Given a system $\mc I$ of constraints on the rank function of a polymatroid with ground set $[N]$, determine if there exists a  polymatroid that satisfies $\mc I$ in $\mathscr P^q(\mb c)$.
\end{itemize}
Note that [E2$_q$] is a special case of  [CLRP$_q$], as a pre-specified rank function $f$ of a polymatroid $(E,f)$ can be interpreted as a collection of $2^N$ linear constraints on the rank function with $N=\vert E\vert$.  As we shall discuss in next section, it is sometimes of interest to find \textit{all} polymatroids satisfying $\mc I$ and belonging to $\mathscr P^q(\mb c)$. We now state CLRP$_q$-EN to be the problem of constructing the members of $\mathscr P(N,(r_l,r_u),K,(s_l,s_u) ) $ up to some notion of equivalence $\equiv$. 
 \begin{itemize}
  \item{[CLRP$_q$-EN]} Given a system $\mc I$ of constraints on the rank function of a polymatroid with ground set $[N]$, list a representative from each equivalence class, under $\equiv$, of polymatroids in $\mathscr P(N,(r_l,r_u),K,(s_l,s_u) ) $ representable over $\mbb F_q$, that satisfy $\mc I$.
\end{itemize}
An algorithm to solve CLRP$_q$-EN can also solve CLRP$_q$-EX, by halting as soon as it finds one (the first) such polymatroid representation. The design of finite-terminating algorithms to solve  CLRP$_q$-EN is the subject of much of this paper. The approach we use is that of using combinatorial generation techniques which are able to construct combinatorial objects satisfying certain desired properties systematically, exhaustively, and efficiently.  

\section{Network Coding, Secret Sharing and CLRP}\label{sec:clrpncss}
We now review the basic terminology related to network coding and secret sharing, showing that creating computer assisted achievability proofs and rate regions in network coding and secret sharing with finite length linear codes can be  posed as variants of CLRP. 
\subsection{Network coding as CLRP}\label{sec:clrpnc}
Multisource multisink network coding over directed acyclic graphs (MSNC) was studied by Yan et. al.\cite{yanmultisourceTIT},
where the authors give an implicit characterization of the rate regions in terms of $\Gamma_N^*$. The version of
network coding problem considered here is \textit{multisource multisink network coding over directed acyclic hypergraphs} (HMSNC),
which was studied recently by Li et. al.\cite{Li_Operators}. HMSNC is a general model that includes as special case the MSNC problem, the
Independent Distributed Source Coding (IDSC) problem, and the index coding (IC) problem \cite{Li_Operators}.

A HMSNC instance is completely described by the tuple $ A=(\mathscr S,\mathscr G,\mathscr T,\mathscr E,\beta)$. It consists of a directed acyclic hypergraph $(\mathscr{V,E})$ where the nodes $\mathscr V = \mathscr S \cup \mathscr T \cup \mathscr G$ can be partitioned into the set of source nodes $\mathscr S$, the set of sink nodes $\mathscr T$ and the set of intermediate nodes $\mathscr G$. $\mathscr E$ is the set of directed hyperedges of the form $(v,\mathscr A)$ where $v\in \mathscr V, \mathscr A\subseteq \mathscr V\setminus \{v\}$.  The source nodes in $\mathscr S$ have no incoming edges and exactly one outgoing edge carrying the source message, the sink nodes $\mathscr T$ have no outgoing edges, and the intermediate nodes $\mathscr G$ have both incoming
and outgoing edges. Any hyperedge $e \in \mathscr E$ then
connects a source or an intermediate node to a subset of non-source nodes, i.e., $e = (i, \mathscr F)$, where $i \in \mathscr S \cup \mathscr G$ and
$\mathscr F\subseteq  (\mathscr G \cup \mathscr T )\setminus \{i\}$. The number of source nodes will be denoted by $\vert \mathscr S\vert=k$, with a source message associated with each node. For convenience, we shall label the source messages on the outgoing hyperedges of source nodes as $1,\hdots,k$. The remaining messages, carried on the rest of the hyperedges, are labeled $k+1,\hdots,\vert \mathscr E\vert$. Thus, we have a total of $N=\vert \mathscr E\vert$ messages. The demand function $\beta: \mathscr T\rightarrow 2^{[k]}$  associates with each sink node a subset of source messages that it desires.  Each message $i\in [N]$ will be associated with 
a discrete random variable $X_i$, giving us the set $\bs X_N$ of message random variables. 

In network coding, it is assumed that the random variables $\{X_i\suchthat i\in [k]\}$ associated with source messages are independent.  Recalling that we are casting network coding rate region problems into the light of CLRPs, the source independence gives rise to the first constraint on the entropy function,  
\begin{equation}
  \mb h_{[k]}=\sum_{i\in \mathscr [k]}\mb h_k.
\end{equation}
For notational convenience, we form the set $\mc L_1$ containing only the above constraint.  For any $v \in \mathscr \mathscr G\cup \mathscr T$ define $\tsf{In}(v)$ to be the set of its incoming messages, and for each $g\in \mathscr G$, define $\tsf{Out}(g)$ to be the set of its outgoing messages.  Every non-source message $i\in [N]\setminus [k]$, originates at some intermediate node $g\in \mathscr G$, and the associated random variable $X_i = f_i(\{X_j,j\in\tsf{In}(g)\})$ is a function of all the input random variables of node $g$. This gives rise to constraints of the following type
\begin{equation}
  \mb h_{\tsf{In}(g)\cup \tsf{Out}(g)}=\mb h_{\tsf{In}(g)},\spc \forall g\in \mathscr G.
\end{equation}
We collect the above constraints into a set $\mc L_2$.  Finally, the demand function requires that each node $t\in \mathscr T$ be able to reconstruct those source messages with indices in $\beta(t)$, which naturally gives rise to the decoding constraints
\begin{equation}
  \mb h_{\tsf{In}(t)\cup \beta(t)}=\mb h_{\tsf{In}(t)},\spc\forall t\in \mathscr T.
\end{equation}
The decoding constraints are collected in a set $\mc L_3$.  We shall denote the collected constraints on the entropy function associated with this network coding problem $A$ as $\mc I_A=\mc L_1\cup \mc L_2\cup \mc L_3$.
\begin{defn}
  A network code for a HMSNC instance $ A$ is a collection of $N$ discrete random variables $\bs X_N$ with joint entropies satisfying the constraints $\mc I_A$.
\end{defn}
Equivalently, a network code is an entropic polymatroid satisfying the constraints $\mc I_A$ on its rank function. 
From the standpoint of the rate region of a network coding problem, all that matters, given the knowledge that the joint entropies satisfy the constraints of the network, is the singleton entropies.  Thus, we will say that a network code $\bs X_N$ \textit{achieves} a rate vector $\mb r= (r_1,\hdots,r_N), r_i\in \mbb Z_{\geq 0}$ if $\mb h_i=r_i,\spc\forall i\in [N]$.  

Let $\mbb R^M$ be the space with subset entropies and rates as co-ordinates, i.e. $M=2^N-1+N$. Aforementioned notion of achieving a rate vector $\mb r$ follows directly from a result of Yan, Yeung, and Zhang \cite{yanmultisourceTIT}, which can be easily generalized to provide the (closure of) the set of all rate vectors achievable for a given network coding problem instance $A$ \cite{Li_Operators}, thereby yielding the capacity region of the network as
\begin{equation}\label{rrexact}
\mc R^*=\tsf{proj}_{\mb r}(\overline{\tsf{con}(\Gamma_N^*\cap \mathscr L)}\cap \mathscr L')
\end{equation}
where $\tsf{con}(\mathscr B)$ is the conic hull of set of vectors $\mathscr B$, $\tsf{proj}_{\mb r}(\mathscr B)$ is the projection onto coordinates $\mb r$ and where 
\begin{equation}
\mathscr L\triangleq \{(\mb h,\mb r)\in \mbb R^M\suchthat \mb h \text{ satisfies $\mc L_1\cup\mc L_2$}\}
\end{equation}
\begin{equation*}
\mathscr L'\triangleq \{(\mb h,\mb r)\in \mbb R^M\suchthat (\mb h \text{ satisfies }\mc L_3)\land (r_i\geq \mb h_{i}, \text{ for each }i\in [N]\setminus [k]) \land (r_i \leq \mb h_{i},\ \text{for each}\ i \in [k] )  \}
\end{equation*}
Under this formulation, the rate region for the network is a convex cone, described by a series of inequalities linking the rates of the sources $r_i, i \in [k]$, with the capacities of the links $r_i, i \in [N] \setminus [k]$.  As $\Gamma_N^*$ is not yet fully characterized for $N\geq 4$, most of the HMSNC rate region characterizations known so far have been found using the method of \textit{sandwich bounds}\cite{DFZMatroidNetworks,Li_Operators}, which is based on substituting polyhedral inner and outer bounds in place of $\Gamma_N^*$ in (\ref{rrexact}). This yields inner and outer bounds $\mc R_{\tsf{in}}$ and $\mc R_{\tsf{out}}$ on $\mc R^*$,
\begin{equation}
\mc R_{\tsf{x}}=\tsf{proj}_{\mb r}(\Gamma_{\tsf{x}}\cap \mathscr L\cap \mathscr L'),\tsf{x}\in\{\tsf{in,out}\} 
\end{equation}
Once we compute $\mc R_{\tsf{in}}$ and $\mc R_{\tsf{out}}$,  if $\mc R_{\tsf{in}}=\mc R_{\tsf{out}}$, we know that $\mc R^*=\mc R_{\tsf{in}}=\mc R_{\tsf{out}}$. 
In the context of this paper, we are particularly interested in rate vectors that are achievable with \textit{linear} network codes.  Given a class $\mathscr P^q(\mb c)$ of linear codes, one can consider the following inner bound on  $\Gamma_N^*$:
\begin{equation}
\Gamma_N^{\mathscr P^q(\mb c)}=\tsf{con}(\{\mb h\in \mbb R^{2^N-1}\suchthat \mb h\in \mathscr P^q(\mb c)\})
\end{equation}
yielding the inner bound  $\mc R_{\mathscr P^q(\mb c)}=\tsf{proj}_{\mb r}(\Gamma_{\mathscr P^q(\mb c)}\cap \mathscr L\cap \mathscr L')$ to the rate region $\mc R^*$.

A linear network code is a representable polymatroid that satisfies constraints $\mc I_A$ on its rank function.  As an aside, note that a linear network code is said to be \textit{scalar} if the associated polymatroid is in fact a matroid.  
From the perspective of achievability proofs for network coding rate regions, and the inner bounds associated with linear codes $\mc R_{\mathscr P^q(\mb c)}$, there are two variants of the constrained linear representability problem that are of interest, an existential one building from CLRP$_q$-EX and an enumerative one building from CLRP$_q$-EN.  The existential variant takes a specified rate vector $\mb r$ and asks whether there is a code over $\mbb F_q$ which achieves it.
\begin{itemize}
  \item{[E3$_q$-EX]} Given a HMSNC instance $A$, determine if there exists  polymatroid of size $N$ representable over $\mbb F_q$ satisfying $\mc I_A$ achieving a rate vector $\mb r$.
  \end{itemize}
Note that E3$_q$-EX is a special case of CLRP$_q$-EX, as we can interpret the requirement to achieve a rate vector $\mb r$ as additional linear constraints $\mc I_{\mb r}=\{\{\mb h_i=r_i\},i\in[N]\}$. Hence, it is an instance of CLRP$_q$-EX with constraints $\mc I=\mc I_A\cup\mc I_{\mb r}$ and class tuple $\mb c=(N,(\sum_i r_i, \sum_i r_i),\tsf{unique}(\mb r),(k,N))$, where $\tsf{unique}(\mb r)$ is the set of unique values in vector $\mb r$.  

The enumerative variant associated with achievability proofs in network coding rate regions aims instead to find, up to isomorphism $\equiv$, all linear codes in a given class that satisfy the constraints of the network.
\begin{itemize}
\item{[E3$_q$-EN]} Given a class tuple $\mb c$ and a HMSNC instance $A$, list a representative from equivalence class of codes (under $\equiv$) yielding joint entropy vectors $\mb h\in \mathscr P^q(\mb c)$ s.t. $\mb h$ satisfies $\mc L_1,\mc L_2$ and $\mc L_3$ associated with $A$.
\end{itemize}
Observe that E3$_q$-EN is likewise a special case of CLRP$_q$-EN associated with the same code class $\mb c$, where again the system of constraints is given by the constraints $\mathscr L$ built from the network coding problem $A$.

Having defined the two problem classes, some comparison and historical discussion is in order.  The existing computation based information theory achievability proofs literature, while thin, is focussed on the former of these two problems E3$_q$-EX.  This matches the situation with the computer aided converse proof literature, which, with the notable exception from our own previous work \cite{jayantchm,Li_Allerton_12,Li_NetCod_13,aptenetcod15,Congduan_MDCS,Li_Operators}, aims to provide proofs verifying a putative given inequality \cite{itip,xitip,TianJSAC433} (i.e. membership testing in the polar $\mc R_{\tsf{out}}^{\circ}$), rather than generating that inequality as part of a full description of an outer bound $\mc R_{\tsf{out}}$ in the first place.  The seminal network coding paper of Koetter and Medard \cite{koetteralgebraic} provided an algebraic formulation of a slight variation of E3$_q$-EX which replaces $\mbb F_q$ with its algebraic closure $\mbb F_q^*$.  Under this Koetter and Medard formulation, one can construct a system of polynomial equations with coefficients in $\mbb F_q$ s.t. non-emptiness of the associated algebraic variety implies the existence of a polymatroid satisfying $\mc I$ that is representable over the algebraic closure of $\mbb F_q$  and vice versa. The problem of determining the emptiness of the algebraic variety associated with these polynomial equations can be solved by computing the Gr\"obner basis \cite{coxidealsbook} of the ideal generated by them in $\mbb F_q[\mb x]$. Shifting back to E3$_q$-EX, i.e. if one is interested in existence of solution over a specific finite field (and not the algebraic closure of it), one can instead compute the Gr\"obner basis in the quotient ring  $\mbb F_q[\mb x]\setminus \langle x_i^q=x_i\suchthat i\in [n]\rangle$.  An important subsequent work of Subramanian and Thangaraj \cite{subrapathgain10} refined the algebraic formulation of Koetter and Medard to switch to using path gain variables instead of variables that represent local coding coefficients.  The benefit of this path gain formulation is that the polynomials contain monomials of degree at most 2, which can have substantial complexity benefits over the original Koetter and Medard formulation in the Gr\"{o}bner basis calculation.  The details of how to adapt this method of Subramanian and Thangaraj to solve E3$_q$-EX are provided in algorithm  \ref{transformhmsnc} in the appendix.  As we will show via computational experiments in \S\ref{sec:example}, the new methods for solving E3$_q$-EX that we will provide in this work, which, in fact, are built from enumerative methods best suited to solving E3$_q$-EN, in some problems enable substantial reduction of runtime relative to the Subramanian and Thangaraj Gr\"{o}bner basis based formulation.  The reduction in runtime appears to be most pronounced when the rate vectors $\mb r$ being tested contain integers greater than one.

More broadly, when one has computed a polyhedral outer bound $\mc R_{\tsf{out}}$ to $\mc R^*$, e.g. through polyhedral projection with the convex hull method \cite{lassezchm,Xu_ISIT_08} as described in our previous work \cite{jayantchm,aptenetcod15}, one can utilize a series of E3$_q$-EX problems to determine if $ \mc R_{\mathscr P^q(\mb c)} = \mc R_{\tsf{out}}$ as follows.  For each extreme ray $\mb r$ of $\mc R_{\tsf{out}}$, one uses E3$_q$-EX to determine if $\mb r$ is achievable with linear codes in the associated class $\mb c$.  If the answer is yes for each of the extreme rays of $\mc R_{\tsf{out}}$, then $\mc R^* = \mc R_{\tsf{out}} = \mc R^* = \mc R_{\mathscr P^q(\mb c)} $.  However, if the answer to E3$_q$-EX is no for one or more of the extreme rays $\mc R_{\tsf{out}}$, then it can be of interest to determine the region $ \mc R_{\mathscr P^q(\mb c)} $ achievable with this class of codes.

This more difficult problem of determining the inequality description of the polyhedral cone $ \mc R_{\mathscr P^q(\mb c)} $ can be solved with an algorithm providing a solution to E3$_q$-EN.  Indeed, denoting the set of vectors forming the solution to E3$_q$-EN as $\mathscr P^q(\mb c,A)$, one can determine $ \mc R_{\mathscr P^q(\mb c)}$ with e.g. the following remaining steps:
\begin{enumerate}
\item Delete all co-ordinates of $\mb h\in \mathscr P^q(\mb c,A)$ other than singletons
\item Append the resulting list of vectors with $-\mb e_i, i \in [k]$ and $\mb e_i, i\in [N] \setminus [k]$, where $\mb e_i$ is the $i$th column of the identity matrix of dimension $N$
\item Compute the inequality description of the conic hull of the resulting set of $N$ dimensional vectors
\end{enumerate} 
The resulting inequalities are the system of inequalities defining the polyhedral cone $ \mc R_{\mathscr P^q(\mb c)}$, and the resulting extreme rays of the conic hull form the optimal codes which can be time-shared to achieve any point in $ \mc R_{\mathscr P^q(\mb c)}$ arbitrarily closely \cite{Congduan_MDCS,Li_Operators}. For the rest of the paper, rate regions will be specified as inequalities amoungst source rate variables $\omega_i,i\in [k]$ and edge rate variables $R_j,j\in [N]\setminus [k]$, whereas the rate vectors will be specified as $(\bs \omega,\mb r)$, where $\bs \omega$ and $\mb r$ are source and edge rate vectors of size $k$ and $N-k$ respectively, to enhance readability.

\subsection{Secret sharing as CLRP}
The next application where the ability to solve CLRP-EX is useful is secret sharing. Secret sharing\cite{Shamir1979ShareSecret,beimelsurvey11} is concerned with the sharing of a secret among a collection of people or entities such that only certain subsets of them can recover the secret. Let $\Delta$ be the set of participants.  We assume that $\vert \Delta\vert=N$, with participants labeled by $[N]$. The secret sharer is called the dealer, and has label 1 while the rest of the participants, called parties have labels in $\{2,\hdots,N\}$. The dealer bears a secret, and gives each party a chunk of information, called a \textit{share}. The subsets of $\Delta\setminus \{1\}$ that are allowed to reconstruct the secret are called the authorized sets. The set of all authorized subsets, called an \textit{access structure}, and denoted by $\Gamma$ is a monotone collection of sets i.e. $\mathscr A\in \Gamma\implies \mathscr B\in \Gamma, \forall \mathscr B\supset \mathscr A$. We associate a random variable $X_1$ with the dealer
and random variables $X_2,\hdots,X_N$ with the parties.
Given an access structure $\Gamma$ containing authorized subsets of
$\{2,\hdots,N\}$, every authorized set of participants must be able to recover the secret, imposing the following constraints on the entropy function.
\begin{equation}\label{sscon1}
\mb h_{\mathscr S\cup\{1\}}=\mb h_{\mathscr S},\spc\forall \mathscr S\in \Gamma 
\end{equation}
Moreover, an un-authorized set must not be able to gain any information about the secret, imposing the following constraints on the entropy function.
\begin{equation}\label{sscon2}
\mb h_1+\mb h_{\mathscr S}=\mb h_{\mathscr S\cup\{1\}},\spc\forall \mathscr S\notin \Gamma
\end{equation}
We denote the collection of constraints specified by \eqref{sscon1} and \eqref{sscon2} as $\mc I_\Gamma$. 
\begin{defn}
A secret sharing scheme (SSS) for a access structure $\Gamma\subseteq 2^{\{2,\hdots,N\}}$ is a collection of $N$ random variables whose joint entropies satisfy the constraints $\mc I_\Gamma$.
\end{defn}
A SSS is linear if it is associated with a representable matroid with ground set size $N$ satisfying $\mc I_\Gamma$. A SSS is said to be multi-linear if it is
associated with a representable polymatroid satisfying $\mc I_\Gamma$.  We can now formulate the achievability problem for secret sharing as follows.
\begin{itemize}
  \item{[E5$_q$]} Given an access structure $\Gamma$, determine if there exists a polymatroid of size $N$ representable over $\mbb F_q$ satisfying $\mc I_{\Gamma}$ with secret size $r_1$ and share sizes $(r_2,\hdots,r_N)$.
  \end{itemize}
Again, [E5$_q$] can be seen as special case of CLRP$_q$ respectively, by interpreting the requirement to have specific secret and share sizes as a collection of constraints $\mc I_{\mb r}=\{\{h_i=r_i\},i\in [N]\}$, where $\mb r=(r_1,\hdots,r_N)$. The class of codes $\mathscr P^q(\mb c)$ is specified by the class tuple $\mb c=(N,(\max_ir_i, \sum_i r_i-1),\tsf{unique}(\mb r),(2,N))$

Secret sharing schemes can be classified into two types: those constructed by putting together other secret sharing schemes (decomposition constructions) and those that are constructed from scratch without using existing schemes (basic constructions). Decomposition constructions of secret sharing schemes have been proposed by Blundo et al.\cite{blundo_graphss95}, Stinson \cite{stinson_graphdecomp94}, van Dijk et al. \cite{vandijk_graph98,vandijk_graph06}. Basic constructions of linear schemes were proposed by Bertilsson et al.\cite{Bertilsson92} and van Dijk \cite{dijk_linearss97}. In \cite{dijk_linearss97}, van Dijk refers to a secret sharing scheme associated with a representable polymatroid as the generalized vector space construction.
He also proposes a backtracking algorithm for determining whether a certain worst case information rate (in context of problem [E5$_q$], $\frac{ \tsf{max}_{i\in\{2,\ldots,N\}} r_i}{r_1}$ is the worst case information rate) can be achieved in a given access structure by means of the generalized vector space construction, thus providing an algorithm to solve a variant of (E5$_q$). Secure network coding (SNC) \cite{cai_securenc} is a generalization of secret sharing.  Again, one can form a variant of CLRP in the context of SNC.

\section{Polymatroid Isomorphism, partial $\mc I$-feasibility, and extension}\label{sec:ingredients}
Algorithms for generating various types of combinatorial objects that are special-cases of polymatroids: graphs, linear codes, matroids and $k$-polymatroids, up to some equivalence relation $\equiv$, have been studied in the literature, see e.g. \cite{blackburnenum,Mayhew2008415,matsumotomatenum,savitsky14,betten2006error}. A common theme in these algorithms is that of "orderly generation". Here, each combinatorial object $X$ has an associated non-negative number "order" $o(X)$ which is the size of the object. This could be the number of edges for graphs, block length for linear codes or the ground set size for matroids/polymatroids. The aim of these algorithms is to list inequivalent objects of order up to some $N\geq 1$. The name orderly generation follows from the fact that these algorithms proceed in an orderly fashion, by constructing inequivalent objects of size $i\leq N$ from the list of inequivalent objects of size $i-1$, recursively in $i$. The key tool used for constructing an object of size $i$ from an object of size $i-1$ is the notion of \textit{augmentation}. For matroids or polymatroids, an example of such operation is extension, which augments a polymatroid with ground set size $i-1$ by adding a new element to the ground set, thereby constructing a polymatroid with ground set size $i$.  
In order to apply this orderly generation technique to solving CLRP$_q$-EN and CLRP$_q$-EX, three decisions must be made.  The remaining subsections of this section describe the answers to these questions in detail, while we give a high level description here.

First, one must specify precisely the combinatorial objects whose generation is being attempted.  In \S\ref{sec:feasibility}, given a specific collection of constraints $\mc I$ on the rank function of a polymatroid of ground set size $N$, we define the concept of linear $\mc I$-polymatroids which is a mathematical formalization of 'polymatroids in $\mathscr P^q(\mb c)$, that satisfy $\mc I$' as mentioned in CLRP$_q$-EN. For a given class of codes $\mathscr P^q(\mb c)$ and constraints $\mc I$ on a polymatroid of ground set size $N$, a  $\mbb F_q$-linear $\mc I$-polymatroid is a pair $(P,\phi)$ where $P$ is a $\mbb F_q$-representable polymatroid in $\mathscr P^q(\mb c)$, having ground set size $N$ and $\phi$ is a bijection $\phi: [N]\rightarrow [N]$. The domain of $\phi$ is understood to be the ground set of $P$ while the range is understood to be the set associated with the set function that $\mc I$ is constraining (we choose to label both these sets with $[N]$). In particular, $\phi$ is a map under which $P$ satisfies $\mc I$. $\phi$ is reminiscent of the network-matroid mapping of Dougherty, Freiling and Zeger\cite{DFZMatroidNetworks}, albeit being set up the other way around (from a polymatroid to a network, when $\mc I$ arises from a network coding instance).  A $\mbb F_q$-representable polymatroid for which such a mapping $\phi$ exists is said to have the property of $\mc I$-feasibility.  In order to embed the constraint of $\mc I$-feasibility into an orderly generation oriented algorithm gradually growing the polymatroid's ground set, we further expand this idea to polymatroids of ground set size $i\leq N$ via the property of partial $\mc I$-feasibllity or $p\mc I$-feasibility, to define linear $p\mc I$-polymatroids, which are $\mbb F_q$-representable polymatroids with ground set size $i\leq N$ for which there exists an injective mapping $\phi:[i]\rightarrow [N]$, under which it satisfies a subset of $\mc I$ associated with $\phi([i])$.  The mapping $\phi$ in this case is called a $p$-map, which is shorthand for partial map.  Noting that the smaller polymatroid created via the deletion of any ground set elements of a $p\mc I$-feasibile polymatroid must also be $p\mc I$-feasible polymatroid, we arrive at the conclusion that the criterion of $p\mc I$-feasibility can be embedded into the orderly generation algorithm.  Namely, if at any time in the extension process we encounter a polymatroid that is not $p \mc I$-feasible, we do not need to compute any extensions of this polymatroid, as they will also not be $p\mc I$-feasible.

In addition to the goal of generating only polymatroids which will ultimately obey the constraints $\mc I$, we are also interested in generating exclusively the \textit{essentially different} ones. This leads to the next of the three decisions which must be made: the notion of equivalence.  The equivalence relation $\equiv$ mentioned while defining CLRP captures this idea.  In \S\ref{sec:polyiso}, we discuss two notions of isomorphism: strong isomorphism and weak isomorphism. Both of these notions have been used in literature to generate combinatorial objects related to polymatroids. Literature related to generation of matroids and 2-polymatroids \cite{blackburnenum,Mayhew2008415,matsumotomatenum} uses strong isomorphism. The literature concerning the generation/classification of linear codes \cite{betten2006error} uses weak isomorphism as it is associated with the semi-linear isometry relation on linear codes of given block-length.  
We use words 'strong' and 'weak' to emphasize the fact that weak isomorphism is a refinement of strong isomorphism, which we elaborate on using example \ref{ex1}.  The benefit of using weak isomorphism for representability polymatroids over strong isormorphism is that it enables group theoretic techniques to provide an algorithm that can exhaustively generate simple $p\mc I$-feasible polymatroids that are distinct under weak isomorphism.  In order to generate representable polymatroids that are distinct under strong isomorphism, while also not necessarily simple, these weakly non-isomorphic can be subjected to further strong isomorphism testing among those pairs whose parameters admit the possibility of strong isomorphism.  Ultimately, we settle on an algorithm which uses both of these notions of isomorphism together, which is implemented in our software \tsf{ITAP}, as described in Section \ref{sec:lowlevel}.

In section \ref{sec:codegentemplate}, we describe a general template for generating all members of a particular class of codes $\mathscr P^q(\mb c)$ up to equivalence relation $\equiv$, which is of interest by itself, as it allows one to construct inner bounds $\Gamma_N^{\mathscr P^q(\mb c)}$ on $\overline{\Gamma_N^*}$. While this template is motivated by algorithms to generate classes of polymatroids more general than linear polymatroids, we use a restrictive definition of polymatroid extension, that preserves $\mbb F_q$-representability, thus avoiding the problem of handling non-representable extensions of representable polymatroids. \S \ref{sec:pmapext} defines a notion of augmentation for $p$-maps, called \textit{$p$-map extension}. This, along with our restrictive notion of polymatroid extension, provides an augmentation operation for $p\mc I$-polymatroids, settling the third key decision of choosing the augmentation operation for designing an orderly generation algorithm. We also provide an algorithmic description of how such augmentation can be performed in practice in \S\ref{sec:pmapext}, along with the details of symmetry exploitation while performing such augmentation in \S\ref{sec:sympmap}.

  

\subsection{$p\mc I$-feasibility, $p\mc I$-polymatroids and $\mc I$-polymatroids}\label{sec:feasibility}
\begin{figure}[h]
\begin{center}
\includegraphics[scale=0.35]{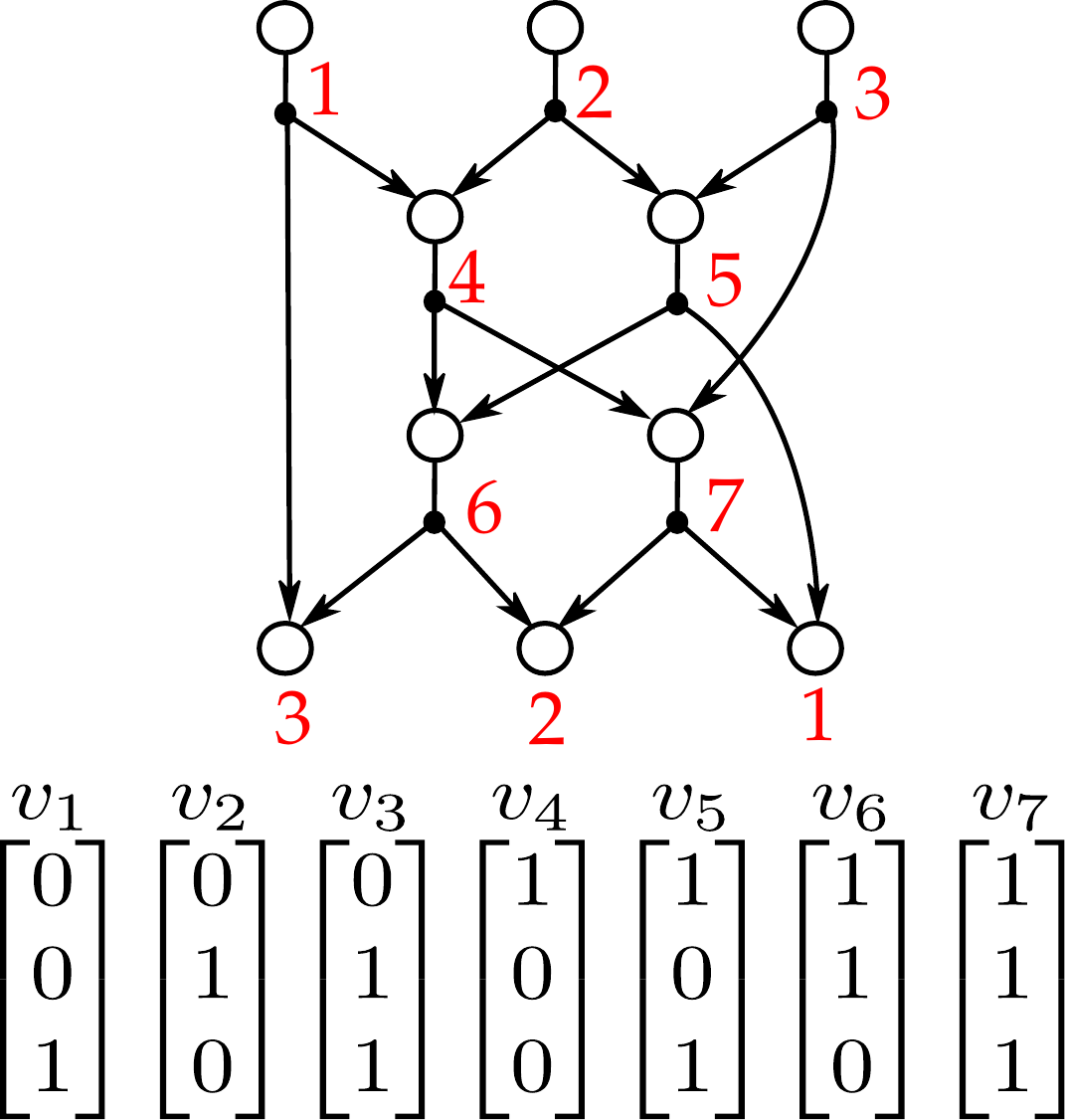}
\end{center}
\caption{(top) the HMSNC instance Fano Network, and (bottom) a representation of the Fano matroid that is an $\mc I$-(poly)matroid with mapping $\phi$ defined as $\{1\mapsto 1,2\mapsto 2,3\mapsto 4,4\mapsto 3,5\mapsto 6, 6\mapsto 5,7\mapsto 7\}$.}
\label{fanoex1}
\end{figure}

\begin{figure}[h]
\begin{center}
\includegraphics[scale=0.5]{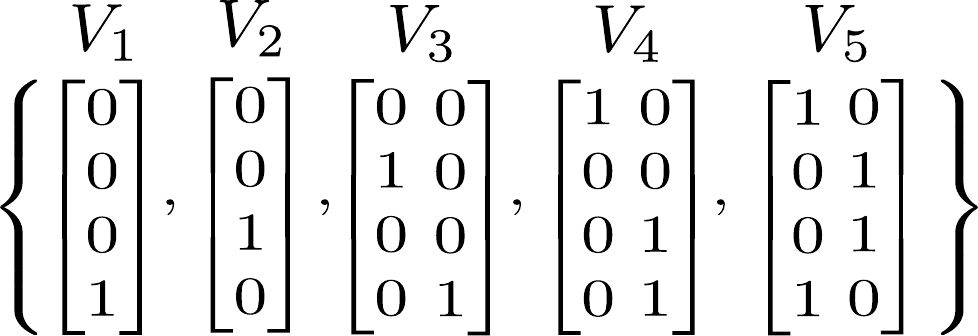}
\end{center}

\caption{An $\mc I$-polymatroid for the collection of constraints $\mc I$ arising from rank vector
$[2,2,4,2,4,4,4,1,2,3,4,3,4,4,4,1,3,3,4,3,4,4,4,2,3,4,4,3,4,4,4]$ with mapping $\phi$ defined as
$\{1\mapsto 4,2\mapsto 5,3\mapsto 1,4\mapsto 3,5\mapsto 2\}$. This polymatroid is an extreme ray of
the cone of linear rank inequalities in $5$ variables\cite{DFZ2009Ineqfor5var}.}
\label{linpoly1}
\end{figure}

 For a collection of constraints $\mc I$ we will first define the property of $p\mc I$-feasibility 
which plays a central role in algorithms discussed later in the paper. For simplicity, we assume that $\mc I$ is presented as a set of linear constraints on the set function defined over $[N]$, with $N$ being called the \textit{size} of $\mc I$. For a polymatroid $P=([i],f)$ with $i\leq N$, we can consider a partial map ($p$-map) that is an injective mapping $\phi$ from from $[i]$ to a $[N]$ that satisfies the relevant part of $\mc I$.  More formally, given a subset $\mathscr X\subseteq [N]$ we denote by $\mc I(\mathscr X)$ the subset of $\mc I$ that contains only those constraints whose involved sets, upon which the set function is evaluated, exclusively to include indices from $\mathscr X$. 
   \begin{defn}
   A polymatroid $([i],f)$ with $i\leq N$ is a $p\mc I$-feasible if there exists an injective mapping $\phi:[i]\rightarrow [N]$ s.t. $f(\phi^{-1}(\cdot))$ satisfies $\mc I(\phi([i]))$. 
   \end{defn}
 We call a $p\mc I$-feasible polymatroid a $p\mc I$-polymatroid, while a $p\mc I$-polymatroid of size $N$ is called a $\mc I$-polymatroid. If $\mc I$ arises from a HMSNC instance, and $([N],f)$ is a matroid, then $\phi^{-1}$ in above definition is known as the \textit{network-matroid} mapping \cite{DFZMatroidNetworks}. 
We shall denote the injective map associated with a $p\mc I$-polymatroid as the $p\mc I$-map and the bijection associated with an $\mc I$-polymatroid as the $\mc I$-map.  Note that the set of all bijections from $[N]$ to itself, denoted as $\Omega$ has size $N!$ while set of all maps from $[i]$ to $\{j_1,\hdots,j_i\}\subseteq [N]$ for $i\leq N$ denoted as $\Omega_p$ contains $\sum_{k=0}^{N-1}\binom{N}{k}(N-k)!$ maps with $\Omega\subseteq \Omega_p$.
Fig. \ref{fanoex1} and Fig.\ref{linpoly1} give examples of an $\mc I$-matroid and an $\mc I$-polymatroid respectively.

Thus, each $p\mc I$-polymatroid is a pair $(P,\phi)$ where $P$ is a polymatroid representation and $\phi$ is a $p$-map. The combinatorial objects we want to systematically generate are $p\mc I$-polymatroids belonging to a particular class of codes $\mathscr P^q(\mb c),\mb c=(N,(r_l,r_u),K,(s_l,s_u))$ up to an equivalence relation $\equiv$.  From orderly generation perspective, each of these has size $N$. The smaller objects of size $i<N$ belong to different classes of linear codes $\mathscr P^q(\mb c_i)$ where $\mb c_i=(i,(r_l,r_u),K,(\max(s_l,i),\min(i,s_u)))$.
\subsection{Notions of polymatroid isomorphism and the equivalence relation $\equiv$}\label{sec:polyiso}
The most natural notion of isomorphism for polymatroids is the following one.
\begin{defn}[Strong Isomorphism] \label{polyiso}
Two polymatroids $P_1=(E_1,f_2)$ and $P_2=(E_2,f_2)$ are said to be isomorphic if there exists a bijection $\phi: E_1\rightarrow E_2$ s.t. $f_1(S)=f_2(\phi(\mathfrak S)),\forall \mathfrak S\subseteq E_1$, denoted as $P_1\cong P_2$.
\end{defn}
 Given two polymatroids $P_1$ and $P_2$, the problem of determining if they are strongly isomorphic has received attention in the literature only in the special case where they are either representable matroids or graphic matroids. Testing if two matroids representable over $\mbb F_q$ are strongly isomorphic is known to be no easier than the graph isomorphism problem \cite{codeeq_complexity}, whose complexity status remains unresolved. There exist necessary conditions for isomorphism, that can be checked in polynomial-time, e.g. for binary matroids \cite{ruditutte}. The problem is somewhat easier if we know beforehand that $\tsf{us}(P_1)=\tsf{us}(P_2)$. Here, we can use the degree vector, as defined below.
\begin{defn}
  The degree vector $(d_1,\hdots,d_{\vert \tsf{us}(P)\vert+1})$ of a polymatroid $P=(E',f')$ with a specified order on $E'$ is an integer vector of size $\vert E\vert+1$ where $(E,f)=\tsf{us}(P)$ with the $i$th entry of the degree vector indicating the size of parallel class of the $i$th smallest non-loop element of $E$ in $P$ for $i \in [\vert E\vert]$, whereas the $\vert E\vert+1$th element specifies the loop degree. 
\end{defn}
An automorphism of a polymatroid is a strong isomorphism from a polymatroid to itself.
\begin{lem}\label{nsiso}
  Let $P_1=(E_1,f_1)$ and $P_2=(E_2,f_2)$ be two polymatroids s.t. $\vert E_1\vert=\vert E_2\vert$ and $\tsf{us}(P_1)=\tsf{us}(P_2)$. Then $P_1$ and $P_2$
  are strongly isomorphic if and only if their associated degree vectors are identical up to an automorphism of $\tsf{us}(P)$.
\end{lem}
To introduce further notions of isomorphism, we first describe the group of semi-linear isometries of $\mbb F_q^r$ and describe its action on the set of all distinct representations of simple linear polymatroids of specified size and rank. Then we explain how this action results in a equivalence relation that is a weakening of the notion of isomorphism in definition \ref{polyiso}. We essentially generalize the notion of semilinear isometry among the generator matrices of projective linear codes over $\mbb F_q$ \cite{betten2006error} to that of weak isomorphism among the distinct representations of simple linear polymatroids.

\begin{defn}
The mapping $\sigma: \mbb F_q^r\rightarrow \mbb F_q^r$ is called semilinear if there exists an automorphism $\alpha$ of $\mbb F_q$ such that for all $\mb u,\mb v\in\mbb F_q^r$ and all $x\in \mbb F_q$, we have
\begin{equation}
\sigma(\mb u+\mb v) =\sigma(\mb u)+\sigma(\mb v),\spc\spc \sigma(x\mb u)=\alpha(x)\sigma(\mb u)
\end{equation}
\end{defn}
A semilinear isometry is a semilinear mapping that maps subspaces to subspaces. The set of all semilinear isometries forms a group known as the general semilinear group.
Let $GL(r,q)$ be the general linear group containing all $r\times r$ invertible matrices over $\mbb F_q$ and let $\tsf{Gal}(q)$ be the Galois group of $\mbb F_q$. $\tsf{Gal}(q)$ is a cyclic group of order $t$ where $q=p^t$ for a prime $p$, generated by the Frobenius automorphism $x\mapsto x^p$. Then the general semilinear group can be defined as follows.
\begin{defn}
The semilinear isometries of $\mbb F_q^N$ form the general semilinear group,
\begin{equation}
\Gamma L(r,q)\triangleq \{(\mbb A,\alpha)\suchthat \mbb A\in GL(r,q),\alpha\in \tsf{Gal}(q)\}
\end{equation} The general semi-linear group $\Gamma L(r,q)$ is the semidirect product $GL(r,q)\rtimes \tsf{Gal}(q)$. The identity element is the pair $(\mbb I_r,\tsf{id})$, where $\tsf{id}$ is the identity element in $\tsf{Gal}(q)$. The multiplication of two elements of $\Gamma L(r,q)$ is given by,
\begin{equation}
(\mbb A_2,\alpha_2)(\mbb A_1,\alpha_1)=(\mbb A\cdot \alpha_2(\mbb A_1),\alpha_2\alpha_1),\spc \forall (\mbb A_1,\alpha_1),(\mbb A_2,\alpha_2)\in \Gamma L(r,q)
\end{equation}
\end{defn}

 $\Gamma L(r,q)$ acts naturally on $\tsf{Gr}_q(r,k)$ as follows
\begin{equation}
\Gamma L(r,q)\times \tsf{Gr}_q(r,k) \rightarrow \tsf{Gr}_q(r,k) : ((\mbb A,\alpha),\langle \mb v_1,\hdots, \mb v_k \rangle)\mapsto \langle \alpha(\mb v_1)\mbb A^t,\hdots, \alpha(\mb v_k)\mbb A^t \rangle
\end{equation}
The subgroup of $\Gamma L(r,q)$ that stabilizes each element of $\tsf{Gr}_q(r,k)$ pointwise is $\{(\mbb A,\tsf{id})\suchthat \mbb A\in \mc Z_r\}$ where $\mc Z_r$ is the center of $GL(r,q)$. This stabilizer is isomorphic to $\mc Z_r$ itself. Hence, $\Gamma L(r,q)$ induces the action of \textit{projective semilinear group} given as,
\begin{equation}
P\Gamma L(r,q)\triangleq \Gamma L(r,q)/\mc Z_r.
\end{equation}
on $\tsf{Gr}_q(r,k)$. The elements of $P\Gamma L(r,q)$ are of the form $(\mbb A\mc Z_r,\alpha)$ for $\mbb A\in GL(r,q)$. Furthermore, it can be expressed as the semidirect product
\begin{equation}
P\Gamma L(r,q)=PGL(r,q)\rtimes \tsf{Gal}(q).
\end{equation}
The identity element is $(\mbb I_r\mc Z_r,\tsf{id})$ where $\tsf{id}$ is the identity element in $\tsf{Gal}(q)$. The multiplication of two elements in $P\Gamma L(r,q)$ is given as
\begin{equation}
(\mbb A_2\mc Z_r,\alpha_2)(\mbb A_1\mc Z_r,\alpha_1) =((\mbb A_2\cdot \alpha_2 (\mbb A_1))\mc Z_r,\alpha_2\alpha_1 )
\end{equation} The inverse element of $(\mbb A\mc Z_r,\alpha)$ is $(\alpha^{-1}(\mbb A))^{-1}\mc Z_r,\alpha^{-1}$.
The action of $P\Gamma L(r,q)$ on $\tsf{Gr}_q(r,k)$ is as follows
\begin{equation}
\gamma_k:P\Gamma L(r,q)\times \tsf{Gr}_q(r,k) \rightarrow \tsf{Gr}_q(r,k) : ((\mbb A\mc Z_r,\alpha),\langle \mb v_1,\hdots, \mb v_k \rangle)\mapsto \langle \alpha(\mb v_1)\mbb A^t\mc Z_r,\hdots, \alpha(\mb v_k)\mbb A^t\mc Z_r \rangle
\end{equation}
Let $k(V)$ denote the dimension of a subspace $V$. The above action can be naturally extended to an action on $\tsf{Gr}_q(r,K)$ which further induces an action on $2^{\tsf{Gr}_q(r,K)}$ as,
\begin{equation}
\gamma':P\Gamma L(r,q)\times 2^{\tsf{Gr}_q(r,K)}\rightarrow 2^{\tsf{Gr}_q(r,K)}:\{V_1,\hdots,V_i\}\mapsto \{\gamma_{k(V_1)}(V_1),\hdots, \gamma_{k(V_i)}(V_i) \}
\end{equation}
for any subset of size $i$ of $\tsf{Gr}_q(r,K)$. We shorten $\gamma'(g,\{V_1,\hdots,V_i\})$ as $\{V_1,\hdots,V_i\}g$, for $g\in P\Gamma L(r,q)$.
Given a collection $K\subseteq  [N]$ of distinct singleton ranks, let $\mathscr S^q(N,r,K)$ be the set of polymatroid representations in $\mathscr P^q(N,r,K,s)$ (described in \S\ref{sub2a}) with $s=N$ i.e. the representations associated with simple polymatroids. One can easily verify that this set is fixed setwise under the above action, as parameters $N,r,K$ remain unchanged. We now define the notion of weak isomorphism.
\begin{defn} (Weak Isomorphism)
Two representations $P_1=\{V_1^1,\hdots V_N^1\},P_2=\{V_1^2,\hdots,V_N^2\}\in \mathscr S_{N,r,K}^q$ are said to be weakly isomorphic if there exists a $g\in P\Gamma L(r,q)$ s.t.
\begin{equation}
\{V_1^1,\hdots V_N^1\}g=\{V_1^2,\hdots,V_N^2\}
\end{equation}
denoted as $P_1\overset{W}{=} P_2$.
\end{defn}

\begin{lem}
Let $P_1,P_2\in \mathscr{S}^q(N,r,K)$ be the representations of two simple $\mbb F_q$-representable polymatroids s.t. $P_1\overset{W}{=} P_2$. Then, $P_1\cong P_2$.
\end{lem}
The opposite of the above statement, however, is not true.   It is possible for even the same $\mbb F_q$-representable polymatroid can have several representations that are weakly non-isomorphic. The following example illustrates this phenomenon.
\begin{exmp}\label{ex1}
Consider the following representations $P_1,P_2\in\mathscr S_{5,5,\{2\}}^2$
\begin{equation}
P_1\triangleq\left\{
\stackrel{\mathlarger{V_1}}{
\begin{bmatrix}
0 & 0\\
0 & 0\\
0 & 0\\
1 & 0\\
0 & 1
\end{bmatrix}},
\stackrel{\mathlarger{V_2}}{
\begin{bmatrix}
0 & 0\\
0 & 0\\
0 & 1\\
0 & 0\\
1 & 0
\end{bmatrix}},
\stackrel{\mathlarger{V_3}}{
\begin{bmatrix}
0 & 0\\
0 & 0\\
0 & 1\\
1 & 0\\
0 & 0
\end{bmatrix}},
\stackrel{\mathlarger{V_4}}{
\begin{bmatrix}
0 & 0\\
0 & 0\\
0 & 1\\
1 & 0\\
1 & 0
\end{bmatrix}},
\stackrel{\mathlarger{V_5}}{
\begin{bmatrix}
0 & 1\\
1 & 0\\
0 & 0\\
0 & 0\\
0 & 0
\end{bmatrix}}
\right\}
\end{equation}
\begin{equation}
P_2\triangleq
\left\{
\stackrel{\mathlarger{V_1'}}{
\begin{bmatrix}
0 & 0\\
0 & 0\\
0 & 0\\
1 & 0\\
0 & 1
\end{bmatrix}},
\stackrel{\mathlarger{V_2'}}{
\begin{bmatrix}
0 & 0\\
0 & 0\\
0 & 1\\
0 & 0\\
1 & 0
\end{bmatrix}},
\stackrel{\mathlarger{V_3'}}{
\begin{bmatrix}
0 & 0\\
0 & 0\\
0 & 1\\
1 & 0\\
0 & 0
\end{bmatrix}},
\stackrel{\mathlarger{V_4'}}{
\begin{bmatrix}
0 & 0\\
0 & 0\\
0 & 1\\
1 & 0\\
1 & 1
\end{bmatrix}},
\stackrel{\mathlarger{V_5'}}{
\begin{bmatrix}
0 & 1\\
1 & 0\\
0 & 0\\
0 & 0\\
0 & 0
\end{bmatrix}}
\right\}
\end{equation}
One can computationally verify that every isomorphism between $P_1$ and $P_2$ must map
$V_5$ to $V_5'$ i.e. fix $V_5$. The subgroup of $PGL(5,2)$ that stabilizes $V_5$ setwise
contains matrices of the form $\begin{bmatrix}
A & \mbb O\\
\mbb O & B
\end{bmatrix}$ where $A\in PGL(2,2)$ and $B\in PGL(3,2)$.
Hence, for $P_1$ and $P_2$ to be weakly isomorphic, we must have $P_1'\overset{W}{=}P_2'$ (eq. \eqref{p1eq},\eqref{p2eq} ) i.e. there must exist a $B\in PGL(3,2)$ that maps $P_1'$ to $P_2'$.
\begin{equation}\label{p1eq}
P_1'\triangleq
\left\{
\stackrel{\mathlarger{W_1}}{
\begin{bmatrix}
0 & 0\\
1 & 0\\
0 & 1
\end{bmatrix}},
\stackrel{\mathlarger{W_2}}{
\begin{bmatrix}
0 & 1\\
0 & 0\\
1 & 0
\end{bmatrix}},
\stackrel{\mathlarger{W_3}}{
\begin{bmatrix}
0 & 1\\
1 & 0\\
0 & 0
\end{bmatrix}},
\stackrel{\mathlarger{W_4}}{
\begin{bmatrix}
0 & 1\\
1 & 0\\
1 & 0
\end{bmatrix}}
\right\}
\end{equation}
\begin{equation}\label{p2eq}
P_2'\triangleq
\left\{
\stackrel{\mathlarger{W_1'}}{
\begin{bmatrix}
0 & 0\\
1 & 0\\
0 & 1
\end{bmatrix}},
\stackrel{\mathlarger{W_2'}}{
\begin{bmatrix}
0 & 1\\
0 & 0\\
1 & 0
\end{bmatrix}},
\stackrel{\mathlarger{W_3'}}{
\begin{bmatrix}
0 & 1\\
1 & 0\\
0 & 0
\end{bmatrix}},
\stackrel{\mathlarger{W_4'}}{
\begin{bmatrix}
0 & 1\\
1 & 0\\
1 & 1
\end{bmatrix}}
\right\}
\end{equation}
If such a $B$ exists, then $W_4^tB=W_i'^t$  for some $i\in[4]$. However, the row reduced echelon form of $W_4^t$ differs from those of $W_1'^t,\hdots,W_4'^t$, contradicting the fact that $B\in PGL(3,2)$. Hence $P_1\not\overset{W}{=} P_2$.
\end{exmp}
Because of the difficulty of isomorphism problems for graphs and their generalizations, some of methods to generate such objects are designed to outright avoid explicit isomorphism testing. The choice of isomorphism relation could be motivated by this issue, as we discuss in \S\ref{sec:lowlevel}.  

\subsection{Polymatroid extension and construction of all members of a class of codes}\label{sec:codegentemplate}
\begin{figure}[h]
\begin{center}
\includegraphics[scale=0.7]{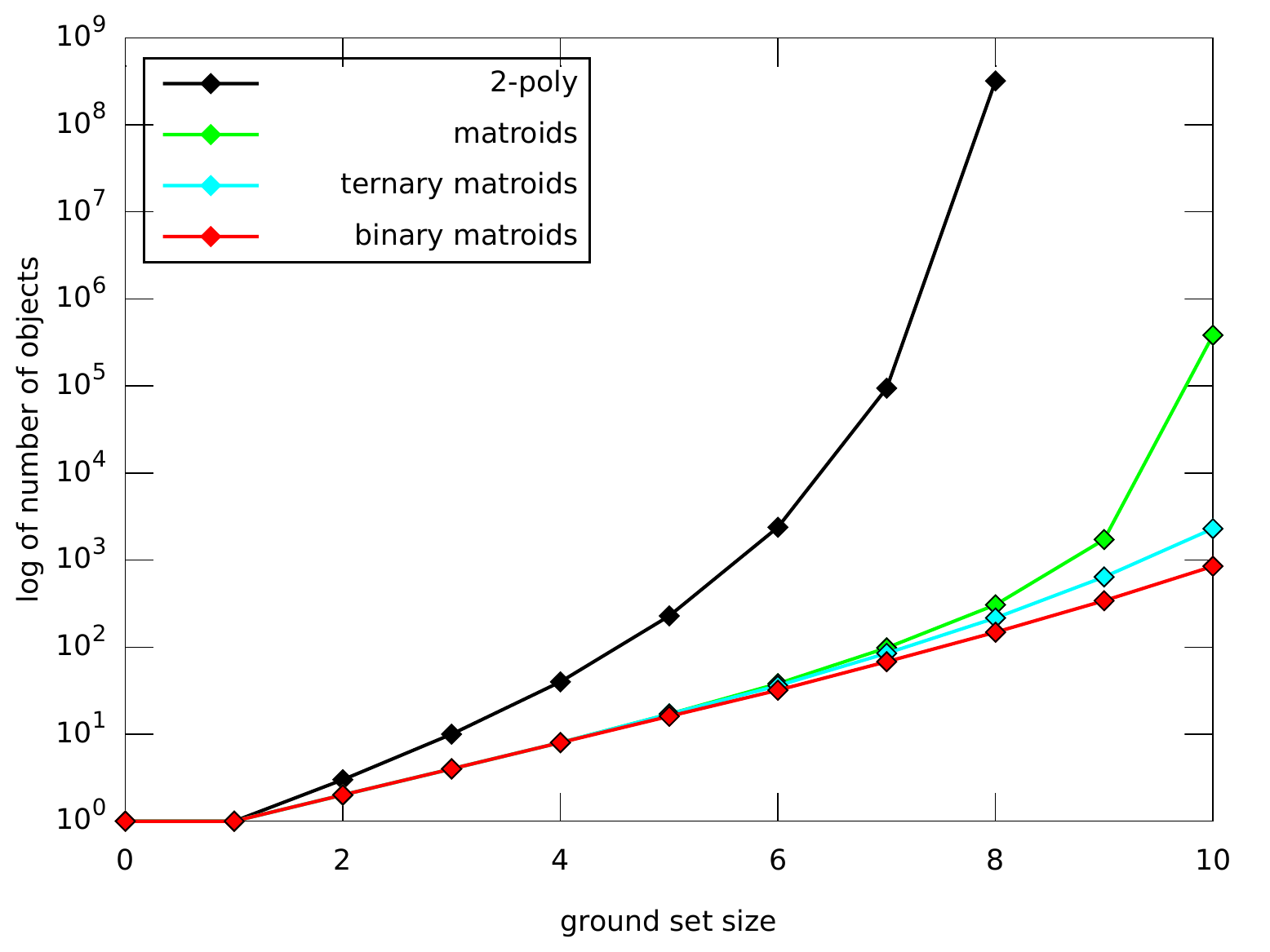}
\end{center}
\caption{Log of number of all (non-isomorphic) $2$-polymatroids \cite{oeis_allpoly}, matroids \cite{oeis_allmatroids}, ternary matroids \cite{oeis_alltern}, and binary matroids \cite{oeis_allbin}, plotted against the ground set size}\label{polygrowth_oeis}
\end{figure}

Polymatroid extension plays a central role in algorithms for exhaustive generation of polymatroids. It provides a means to construct polymatroids on ground set size $i+1$ from polymatroids on ground set size $i$. The following is a general definition of polymatroid extension which applies to arbitrary (not necessarily integer or representable) polymatroids. 
\begin{defn}\label{generalextdef}
  A polymatroid $(E',f')$ is said to be an extension of polymatroid $(E,f)$ if
  $E\subseteq E'$ and $f'(\mathfrak S)=f(\mathfrak S)\spc \forall \mathfrak S\subseteq E$. Furthermore, if
  $\vert E'\setminus E\vert=t$ then $(E',f')$ is said to be a $t$-extension of
  $(E,f)$.
\end{defn}
If $t$ is $1$ and $E'\setminus E=\{e\}$, then we say that $(E',f')$ is a $1$-extension of $(E,f)$ by an element $e$. 
The theory of unique 1-extensions for the special case of matroids was developed by Crapo \cite{craposee} where he provided a method to construct all unique $1$-extensions of a matroid. This method, along with explicit isomorphism testing was used by Blackburn et al.\cite{blackburnenum} to construct all non-isomorphic matroids on up to 8 elements. More recently Mayhew and Royle \cite{Mayhew2008415} constructed all matroids on 9 elements using the same method. Matsumoto et al. \cite{matsumotomatenum} extended Crapo's theory to avoid constructing isomorphic single element extensions and partially constructed matroids on 10 elements.  Savitsky \cite{savitsky14} generalized Crapo's method of producing unique $1$-extensions of matroids to integer $k$-polymatroids, which are integer polymatroids with the rank of every singleton bounded above by $k$ and generated a catalog of $2$-polymatroids on up to $7$ elements. Unfortunately, techniques described by Crapo et. al. \cite{craposee}  for matroids and Savitsky for  \cite{savitsky14}  for $k$-polymatroids are far too general for our purpose. Matroid extensions followed by representability checking style algorithm has been proposed in e.g. \cite{Apteisit14}, but has met with limited practical success for several reasons.  Firstly, if $(E,f)$ is a representable polymatroid, the extensions produced are not guaranteed to be representable, which  means we must employ additional computation to find and weed out the ones that are non- representable.   Secondly, these methods are only developed for polymatroids with specific sets of singleton ranks i.e. $K_P= \{0,1\}$ in case of Crapo et al.'s work and $K_P= \{0,1,\hdots,k\}$ for some $k\in \mbb N$ in case of Savitsky's work.  The third factor that discourages one from using such overly general techniques is the sheer rate at which number of general polymatroids or matroids grows as compared to their representable counterparts. For instance, in fig. \ref{polygrowth_oeis} we can see the growth of number of all (non-isomorphic, integer valued) 2-polymatroids, all matroids, all ternary representable matroids and all binary representable matroids. 

Bearing this in mind, we instead aim to enumerate representable polymatroids directly exclusively, and hence, we define a notion of extension for a polymatroid which we choose to use over the more general notion in def. \ref{generalextdef}, as it conveniently refers directly to the representation of the polymatroid.
\begin{defn} Let $P$ be a polymatroid in $\mathscr P^q(i,(r,r),K,(0,i))$ represented as $\{V_1,\hdots,V_i\}$. Then, a $1$-extension of $P$ is any polymatroid  $P'$ represeted as $\{V_1,\hdots,V_i,V_{i+1}\}$ where $V_{i+1}\in \tsf{Gr}_q(r,K)$.
\end{defn}
Note that the above definition augments a polymatroid made up of a multiset of subspaces of $\mbb F_q^r$ by adding another subspace of $\mbb F_q^r$ to the multiset, thus keeping the underlying vector space dimension $r$ as a constant.  We further classify $1$-extensions into two types: simple $1$-extensions and a non-simple $1$-extensions. A $1$-extension is simple if $\vert \tsf{us}(P')\vert =\vert \tsf{us}(P)\vert +1$  and non-simple otherwise. For an extension to be simple, $V_{i+1}$ must be distinct from each of $V_1,\hdots,V_i$. On the other hand, an extension is non-simple if and only if $V_{i+1}$ is a copy of one of $V_1,\hdots,V_i$ or it is an empty subspace with $f(i+1)=0$. 

Now we describe our basic strategy for constructing all polymatroids in $\mathscr P^q(N,(r_l,r_u),K,(s_l,s_u))$ up to some notion of equivalence $\equiv$. A basic assumption about $\equiv$ is that for polymatroids belonging to the same equivalence class under $\equiv$, the parameters $N,r,K$ and $\vert \tsf{us}(P)\vert$ are identical, which is indeed the case with both strong and weak isomorphism relations discussed in \ref{sec:polyiso}. We also assume that we have access to procedures $\tsf{se}(\cdot)$ and $\tsf{nse}(\cdot)$ which take as input a list of $\equiv$-inequivalent polymatroid representations and produce a list of $\equiv$-inequivalent simple and non-simple $1$-extensions respectively. We defer the discussion of low-level details of how these procedures work in practice to the next section, where we point out several techniques in literature that can be used to implement such procedures. 
\begin{figure}[h]
\begin{center}
\includegraphics[scale=0.7]{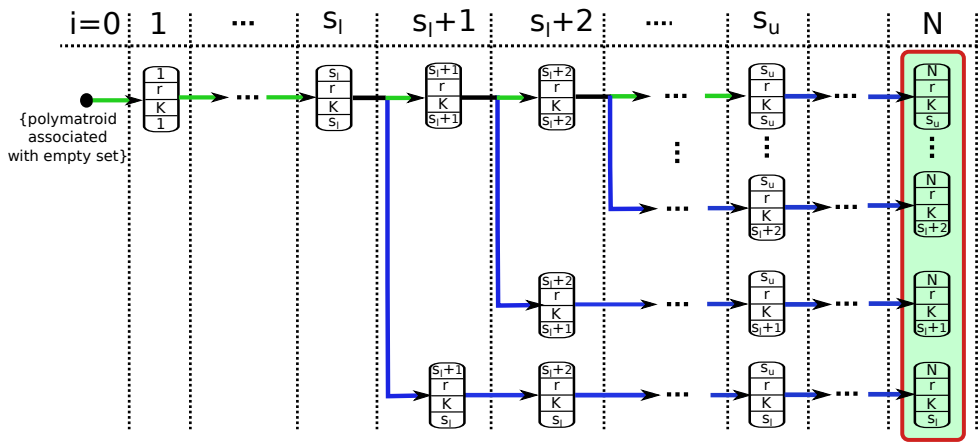}
\end{center}
\caption{Construction of $\equiv$-inequivalent polymatroid representations belonging to the class $\mathscr P^q(N,(r_l,r_u),K,(s_l,s_u))$, for some $r_l\leq r\leq r_u$. The green arrows indicate the use of simple extensions $\tsf{se}(\cdot)$ while blue arrows indicate the use of non-simple extensions $\tsf{nse}(\cdot)$. Each box itself corresponds to a particular class of codes.The parameters specified from top to bottom are: 1) size or length $i$, dimension $r$ of vector space over $\mbb F_q$ whose subspaces are used to build each polymatroid in the class, 3) set $K$ of distinct singleton ranks, and 4) $|\tsf{us}(P)|$ for each polymatroid $P$ in the class.}\label{codecon_strategy}
\end{figure}
Fig. \ref{codecon_strategy} describes how one can use procedures $\tsf{se}(\cdot)$ and $\tsf{nse}(\cdot)$ to construct all members of $\mathscr P^q(N,(r,r),K,(s_l,s_u))$ up to an equivalence relation $\equiv$. This strategy can be used repeatedly for different values of $r$ s.t. $r_l\leq r\leq r_u$ to construct all members of $\mathscr P^q(N,(r_l,r_u),K,(s_l,s_u))$ up to $\equiv$. While our goal in this paper is to solve variants of CLRP$q$, the strategy described in Fig. \ref{codecon_strategy} can be used to construct inner bounds $\Gamma_N^{\mathscr P^q(\mb c)}$, which might be of interest in their own right.
\subsection{An augmentation operation for $p\mc I$-polymatroids}\label{sec:pmapext}
We now describe the construction of all $p\mc I$-polymatroids of size $i+1$ from non-isomorphic $p\mc I$-polymatroids of size $i<N$ by combining linear $1$-extension of $p\mc I$-polymatroids and extension of the associated partial maps. In the background setting provided by the strategy described in Fig. \ref{codecon_strategy}, to generate linear $p\mc I$-polymatroids, the basic problem we must be able to solve is that of determining if a given polymatroid is a $p\mc I$-polymatroid:  
\begin{itemize}
\item{[X1]} For a collection of constraints $\mc I$ of size $N$, given a polymatroid $P=([i],f)$, $i\leq N$, determine if $P$ is a $p\mc I$-polymatroid.
\end{itemize}
In order to settle [X1], we must look for a $p$-map $\phi$ under which $P$ satisfies $\mc I$. If we are successful in finding a certificate \pmap\spc $\phi$ showing that $P$ is a $p\mc I$-polymatroid, we decide to keep $P$. Otherwise we reject $P'$ and all its $t$-extensions from contention.

Problem [X1] can be solved by using a backtracking approach.  To see how, consider set of all possible $p$-maps $\Omega_p$, ordered using the lexicographic order described as follows. Let $\delta$ be the map $\{1\mapsto i_1,\hdots,{\vert\delta\vert}\mapsto i_{\vert\delta\vert}\}$ and $\gamma$ be the map $\{1\mapsto j_1,\hdots,{\vert\gamma\vert}\mapsto j_{\vert\gamma\vert}\}$ where ${\vert\delta\vert}$ and ${\vert\gamma\vert}$ denote the size of the domain of $\delta$ and $\gamma$ respectively with $i_k\in [N],\forall k\in[\vert \delta\vert]$ and $j_k\in [N],\forall k\in[\vert \gamma\vert]$. Denote the ordinary lexicographic (dictionary) order on tuples of length $\leq N$ with elements from $[N]$ as $\overset{L}{<}$, then this ordering is directly extendable to partial maps as follows.
\begin{defn}
Let $\delta,\gamma\in \Omega_p$ be distinct $p\mc I$-maps. Then $\delta$ is smaller than $\gamma$ if $(i_t)_{t=1}^{\vert\delta\vert}\overset{L}{<} (j_t)_{t=1}^{\vert\gamma\vert}$.
\end{defn}
We now define the $p$-map tree which is a directed tree whose vertices are $p$-maps and the edges are defined using the notion of an \textit{extension} of a $p$-map.
\begin{defn}
A \pmap\spc$\delta':[i+1]\rightarrow [N]$ is said to be an extension of a \pmap\spc $\delta:[i]\rightarrow [N]$ if $i< N$ and $\delta(k)=\delta'(k),\spc \forall k\in [i]$.
\end{defn}
We say that the \pmap\spc $\delta$ in the above definition is a deletion of \pmap\spc $\delta'$.
The \pmap\spc tree of order $N$ is a directed graph $T_N=(V_N,E_N)$ where $V_n$ is the set of all
$p$-maps and $(u,v)\in E$ if $v$ is an extension of $u$.
\begin{figure}[h]
\begin{center}
\includegraphics[scale=0.5]{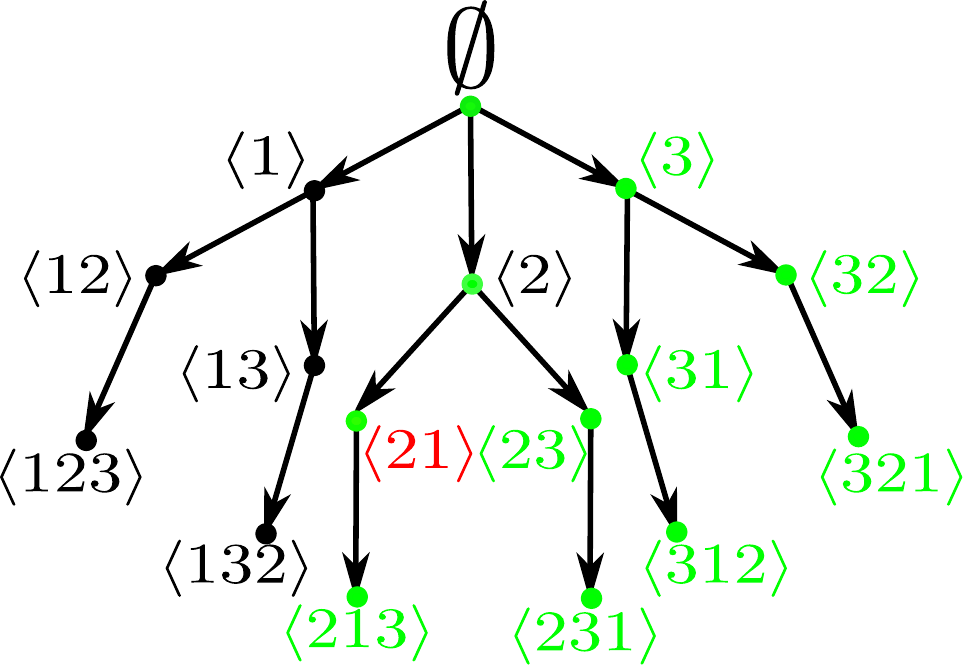}
\end{center}
\caption{The \pmap\spc tree $T_3$ with the subtree $T_3^{>\langle21\rangle}$. For a collection of constraints $\mc I$ of size 3, let a size 2 polymatroid $(E,f)$ be a $p\mc I$-polymatroid with $p\mc I$-map $\{1\mapsto2,2\mapsto1\}$ (shown in red), and let $(E',f')$ be its $1$-extension. Then we need only traverse the vertices shown in green in worst case to determine if $(E',f')$ is also a $p\mc I$-polymatroid.}
\label{lextree}
\end{figure}
Note that \pmap\spc extension and deletion provide a means to traverse the \pmap\spc tree in lexicographic order. Furthermore, given a vertex $u$ (a \pmap) in the tree, one can resume the traversal to produce all $p$-maps that are lexicographically greater than $u$.  Given a \pmap\spc $\delta=\langle i_1,\hdots, i_k\rangle$ at depth $k<N$ in the \pmap\spc tree $T_N$, its immediate descendents can be computed as $\langle i_1,\hdots,i_k,j\rangle$ where $j\in [N]\setminus\{i_1,\hdots,i_k\}$ and can be visited in lexicographical order. The parent of $\delta$ can be obtained by deleting $i_k$. Thus, we need not explicitly store the \pmap\spc tree in order to traverse it.

One merit of considering the lexicographic ordering of $p$-maps and thinking of the collection of all such p-maps as a tree ordered under this ordering is that it allows the problem of finding a $p$-map, or lack of existence thereof, to be posed as a tree search algorithm.  One way to solve [X1] is to traverse $T_N$ in a depth first manner to depth $i$, while testing constraints $\mc I(\delta([j+1]))-\mc I(\delta([j]))$ at each node associated with a \pmap\spc $\delta$ at depth $j$, $j\leq i$.  In this instance, what is meant by depth first is that one extends a partial map adding mapped elements one by one until is it no longer possible to satisfy the constraints $\mc I$ using it, i.e., until all 1 extensions of the partial map no longer obey the constraints.  At that time, a backtracking step deletes the most recent extension, and the partial map extension process continues. If ever a complete traversal to depth $i$ succeeds, then a $p$-map of size $i$ has been found, and provides a certificate showing that [X1] has been solved and that $P=([i],f)$ is a $p\mc I$-polymatroid.  Conversely, if the entire tree is exhausted, i.e. the algorithm terminates with no more backtracks or extensions possible, then [X1] has been answered in the negative.
Of course, our primary interest in this manuscript is the construction of polymatroids with rank functions obeying a series of linear constraints through an extension process i.e. we would like be able to use the strategy in Fig. \ref{codecon_strategy} while at the same time maintaining only those polymatroids that are $p\mc I$-feasible.  The tree structure of the set of $p$-maps is also beneficial in this extension process.  In particular, consider a $p\mc I$-polymatroid $P=([i],f)$ of size $i<N$ with a \pmap\spc $\phi$, and let $P'=([i+1],f)$ be a $1$-extension of $P$.  Then, the $p\mc I$-polymatroid extension process must solve the following problem at each step.
\begin{itemize}
\item{[X2]} Determine if $P'=([i+1],f)$, the $1$-extension of a $p\mc I$-polymatroid $P=([i],f)$ with lexicographically minimum $p$-map $\phi$, is a $p\mc I$-polymatroid, and if so, find its lexicographically minimum $p$-map.
\end{itemize}
That is, we must determine if there exists a \pmap\spc $\phi'$ for $P'$ s.t. it is a $p\mc I$-polymatroid. Given $\phi$, we can use the following lemma to traverse only a subtree of $T_N$ for determining if $P'$ is a $p\mc I$-polymatroid.
\begin{lem}
Let $P$ be a $p\mc I$-polymatroid with $\phi$ being the lexicographically smallest $\pmap$ associated with it and let $P'$ be a $1$-extension of $P$. If $P'$ is a $p\mc I$-polymatroid with $\phi'$ being the $p$-map associated with it, then $\phi'\overset{L}{>}\phi$.
\end{lem}
\noindent This lemma shows that we can determine whether the polymatroid 1-extension $P'$ is a $p\mc I$-polymatroid by traversing the \pmap\spc tree in depth-first fashion, resuming the tree traversal from $\phi$. Let $V_i^{>\delta}$ be the set of all $p$-maps that are lexicographically greater than $\delta$ and $\hat{V}_i^\delta$ be the set of all ancestors of $\delta$. Denote by $T_i(\delta)$ to be the subgraph of $T_i$ induced by vertices $V_i^>\cup \hat{V}_i^\delta$ and the vertices. Hence, it suffices to traverse $T_i(\delta)$ to settle [X2].

\subsection{Exploiting symmetry when augmenting $p\mc I$-polymatroids}\label{sec:sympmap}
In many instances, depending on the low level techniques used for implementing procedures $\tsf{se}(\cdot)$ and $\tsf{nse}(\cdot)$, either a symmetry group of the polymatroid $P=([i],f)$ at the current step of the polymatroid extension process,  or a symmetry group of the constraints $\mc I$ or both might be known.   Knowledge of these groups can further reduce the amount of computation required in searching for $p$-maps in problem [X2].   In particular, we can assume that the symmetries of $P=([i],f)$ are specified as a group $A\leq S_i$ where $S_i$ is the group of all permutations of $[i]$: so that for each permutation $a \in A$ and for each set $\mathscr E \subseteq [i]$, $f(a(\mathscr E)) = f(\mathscr{E})$.  Similarly, the symmetries of $\mc I$ are provided as a group $B\leq S_N$ where $S_N$ is the group of all permutations of $[N]$: for each element $b \in B$, if $([N],f')$ is $p\mc I$-polymatroid, then so is $([N],f'\circ b)$.  Together, as each putative partial map for $P=([i],f)$ is an injective map $\phi: [i] \rightarrow [N]$, the direct product $A\times B$ acts on a putative partial map $\phi$ via $((a,b),\phi) \mapsto b (\phi(a(\cdot)))$.  When attempting to search for a $p$-map, then, one only needs to consider single representatives from the equivalence class created on the $p$-maps under this group action.

\begin{figure}[h]
\begin{center}
\includegraphics[scale=0.7]{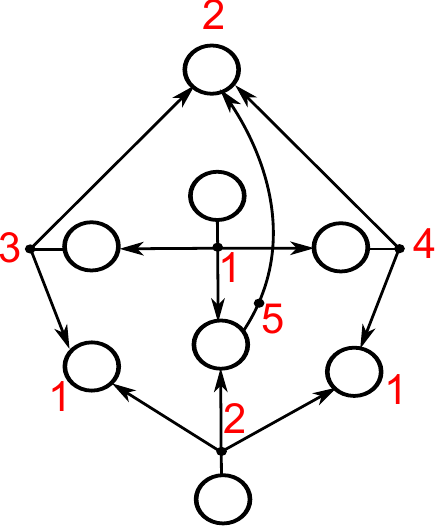}
\end{center}
\caption{A HMSNC instance with $N=5$. The symmetry group $A$ of this instance is of order $2$ generated by $\{(3,4)\}$.}
\label{net22}
\end{figure}

\begin{figure}[h]
\begin{center}
\includegraphics[width=0.9\textwidth]{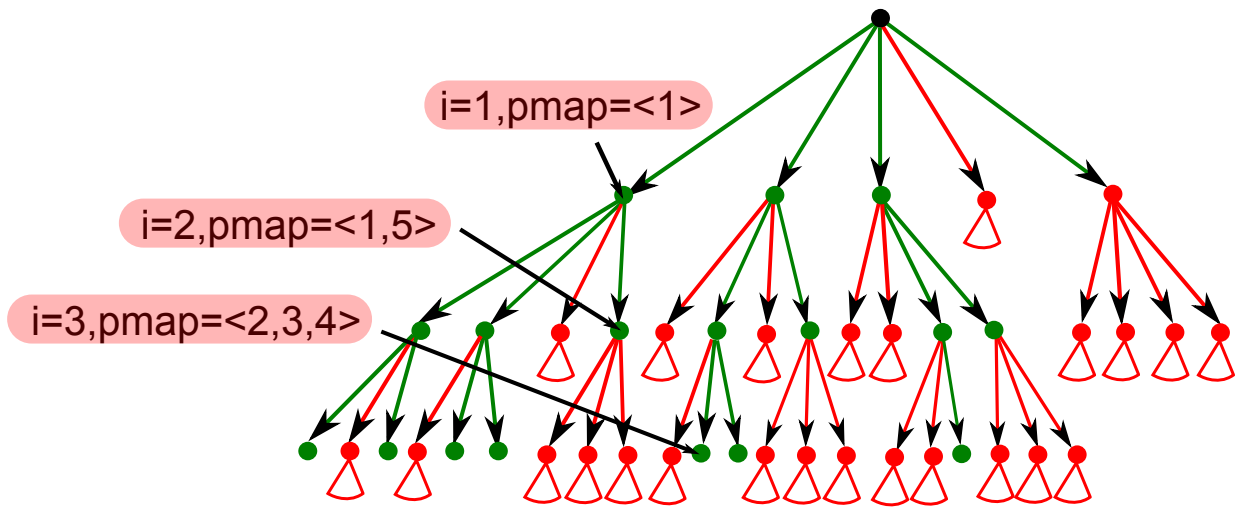}
\end{center}
\caption{The nested $p$-map trees $T_i$, $i\leq 3$ for HMSNC instance in fig. \ref{net22}.  The subtrees $T_i^*$  are shown in green. Each $T_i^*$ is the subgraph of $T_i$ induced by vertices associated $p$-maps that are with lexicographically smallest in their respective orbits under the the action of $A\times B$ where B is the trivial group for $i=1$, and is generated by $\{(1,2)\}$,$\{(1,2),(1,3,2)\}$ for $i=2,3$ respectively. The vertices at every level are drawn in lexicographic order (ascending from left to right).}
\label{lextreeitself}
\end{figure}

Formally, the associated problem can be stated as follows.
\begin{itemize}
\item{[X3]} Given symmetry groups $A,B$ determine if $P$ is a $p\mc I$-polymatroid.
\end{itemize}
At depth $j\leq i$ in $T_N$, denote by $A_{[j]}$ the subgroup of $A$ that stabilizes $[j]$ set-wise. We now consider the action of the direct product $G_{[j]}\triangleq A_{[j]}\times B\leq A\times B$ on $V_j$ which is the set of vertices of $T_N$ at depth $j\leq i$. 
The composition in $G_{[j]}$ is denoted as $((a_1,b_1)*(a_2,b_2))=(a_1a_2,b_2,b_2)$. One can see that $((a_1a_2,b_2,b_2)\delta)(x)=b_1b_2\delta (x(a_1a_2))=b_1(b_2\delta ((xa_1)a_2))=b_1((a_2,b_2)f)(xa_1)=(a_1,b_2)(((a_2,b_2)f)(x))$ for any $x\in [j],j\leq i$ where $[j]$ is the domain of \pmap\spc $\delta$. Let $V_j^*$ be the transversal of orbits in $V_j$ under the aforementioned action formed by choosing the lexicographically smallest \pmap\spc from each orbit. Let $T_i^*$ be the subgraph of $T_i$ induced by $\cup_{j\leq i} V_j^*$.
\begin{lem}
It suffices to traverse $T_i^*$ instead of $T_i$ to determine if $P=([i],f)$ is a $p\mc I$-polymatroid.
\end{lem}
\begin{exmp}
Consider the HMSNC instance shown in fig. \ref{net22}. Let the associated constraints be $\mc I$. This network has a symmetry group of order $2$ generated by permutations \{(3,4)\}.  The polymatroid $\{V_1,V_2,V_3\}$ in \eqref{net22poly} is a $p\mc I$-polymatroid obtained via extending $p\mc I$-polymatroids $P_1=\{V_1\}$ and $P_2=\{V_1,V_2\}$, in that order. Fig. \ref{lextreeitself} shows how the knowledge of network symmetry group, along with symmetries of the polymatroids $P_i$ canbe used to traverse only a subset of the verticces of the $p$-map tree when determining their $p\mc I$ feasibility. 
\begin{equation}\label{net22poly}
P_3\triangleq
\left\{
\stackrel{\mathlarger{V_1}}{
\begin{bmatrix}
0 & 0\\
1 & 0\\
0 & 1
\end{bmatrix}},
\stackrel{\mathlarger{V_2}}{
\begin{bmatrix}
1 & 0\\
0 & 0\\
0 & 1
\end{bmatrix}},
\stackrel{\mathlarger{V_3}}{
\begin{bmatrix}
1 & 0\\
0 & 1\\
0 & 0
\end{bmatrix}}
\right\}
\end{equation}
\end{exmp}
Problems [X1]-[X3] can be combined to form problem [X4] below:
\begin{itemize}
\item{[X4]} Determine if $P'=([i+1],f)$ is a $p\mc I$-polymatroid given symmetry groups $A,B$ and the $p$-map $\phi$ associated with $p\mc I$-polymatroid $P=([i],f)$ where $P'$ is a $1$-extension of $P$.
\end{itemize}
The above problem can be solved by combining the approaches to [X1]-[X3] i.e. it suffices to traverse $T_{i+1}^*(\phi)$, where, using induction on $i$, we assume that $\phi$ is part of $T_{i+1}^*$.

The functionality of $p$-map extension, for generation of $\mc I$-polymatroids, is provided by the procedure $\text{extend\_pmap}(P',\tsf{pc},G)$ that accepts a $1$-extension $P'$ of a $p\mc I$-polymatroid $P$, the $p$-map $\tsf{pc}$ and a group $G$ of symmetries of $\tsf{us}(P')$. Note that symmetries of $P'$ can be directly deduced from those of $\tsf{us}(P')$. If $P'$ is indeed a $p\mc I$-polymatroid, then it returns a $p$-map $\tsf{c}$ associated with $P'$ s.t. $\tsf{c}\overset{L}{>} \tsf{pc}$ which serves as a certificate. Otherwise, it returns empty $p$-map, denoted as $\phi_{\tsf{null}}$. Note that \pmap\spc $\tsf{c}$ produced in this manner will itself be the lexicographically smallest such map associated with $P'$.

\section{An algorithm for solving CLRP$_q$-EN}\label{sec:polyext}
In this section, we build on various concepts described in previous section, to provide the high level description of an algorithm to solve CLRP$_q$-EN via exhaustive generation of $p\mc I$-polymatroids up to equivalence relation $\equiv$. The description is generic enough so that $\equiv$ can be interpreted as either strong or weak isomorphism.  We also discuss how procedures $\tsf{se}(\cdot)$ and $\tsf{nse}(\cdot)$ could be implemented in \S\ref{sec:lowlevel}. 
\subsection{High-level description of the algorithm} 
In \S\ref{sec:codegentemplate} we discussed how to generate all members of a class of codes $\mathscr P^q(\mb c)$ up to an equivalence relation $\equiv$. We intend to use fig. \ref{codecon_strategy} as a template for generating $p\mc I$-polymatroids. To that end, we establish some facts about the property of $p\mc I$-feasibility, that allow us to restrict the strategy in fig \ref{codecon_strategy} to $p\mc I$-polymatroids only. The two main facts that help us achieve this are the \textit{inheritedness} and \textit{isomorph-invariance} of the property of $p\mc I$-feasibility. 
\begin{defn}
A property $\chi$ of polymatroids is said to be inherited if a polymatroid $P$ has $\chi$ then all polymatroids obtained from $\mc P$ via deletion of ground set elements also have $\chi$.
\end{defn} 
Let $P$ be a $p\mc I$-polymatroid with associated $p$-map $\phi$. One can form a $p$-map for polymatroid $P'$ obtained by deletion from $P$ by simply deleting mappings associated with deleted ground set elements. Such a $p$-map serves as a certificate of $p\mc I$-feasibility of $P'$, thus showing that property of $p\mc I$-feasibility is inherited. As an implication, every $p\mc I$polymatroid of size $i$ can be obtained via polymatroid extension and $p$-map extension from some $p\mc I$-polymatroid on ground set of size $i-1$. 
\begin{defn}
A property $\chi$ of polymatroids is said to be $\equiv$-invariant wrt equivalence relation $\equiv$, if a polymatroid $P$ has $\chi$, then all polymatroids equivalent to it under $\chi$ also have $\chi$.
\end{defn}
For both equivalence relations $\cong$ and $\overset{W}{=}$ discussed in this work, members of the same equivalence class have the same rank function up to a permutation of the ground set. Hence, $p\mc I$-feasibility is both $\cong$-invariant and $\overset{W}{=}$-invariant, as given a certificate $\phi$ of $p\mc I$-feasibility of one member of the equivalence class, we can construct such a certificate for every member of the said equivalence class, simply by applying an appropriate permutation to the domain of $\phi$. 
\begin{figure}[h]
\begin{center}
\includegraphics[width=\textwidth]{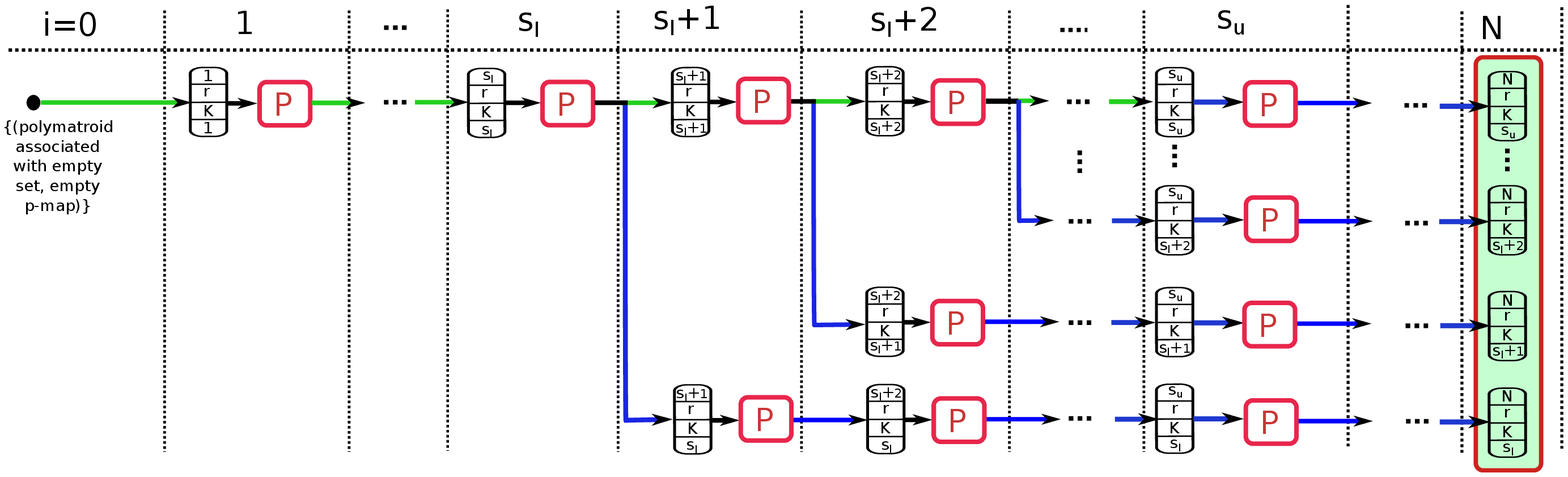}
\end{center}
\caption{Construction of $\equiv$-inequivalent $p\mc I$-polymatroids belonging to the class $\mathscr P^q(N,(r_l,r_u),K,(s_l,s_u))$, for some $r_l\leq r\leq r_u$. The green arrows indicate the use of simple extensions $\tsf{se}(\cdot)$ while blue arrows indicate the use of non-simple extensions $\tsf{nse}(\cdot)$. Each box itself corresponds to polymatroids belonging a particular class of codes. The parameters specified from top to bottom are: 1) size or length $i$, dimension $r$ of vector space over $\mbb F_q$ whose subspaces are used to build each polymatroid in the class, 3) set $K$ of distinct singleton ranks, and 4) $\tsf{us}(P)$ for each polymatroid $P$ in the class. The red box labeled P corresponds to procedure $\tsf{extend}\_\tsf{pmap}(\cdot)$ that extends $p$-maps of the parents of polymatroids to find certificates of $p\mc I$-feasibility while filters out polymatroids for which no such extension exists.}\label{clrp_codecon_strategy}
\end{figure}
Fig. \ref{clrp_codecon_strategy} describes our algorithm to solve CLRP$_q$-EN. This is obtained essentially by adding the functionality of $p$-map extension to the strategy described in fig. \ref{codecon_strategy}. Any polymatroids that are found to be not $p\mc I$-feasible are filtered out, a convenient feature is made possible by inheriteness and isomorph-invariance of the property of $p\mc I$-feasibility.

\subsection{Low-level details: simple and non-simple polymatroid extensions}\label{sec:lowlevel}
This section is concerned with the implementation of procedures $\tsf{se}(\cdot)$ and $\tsf{nse}(\cdot)$. The two main aspects one must pay attention to are: a) the construction of \textit{distinct} $1$-extensions of a polymatroid and b) isomorph rejection. 
In case of $\tsf{se}(\cdot)$, we are given a list of $\equiv$-inequivalent polymatroids belonging to a class $\mathscr P^q(i-1,(r,r),K,(s,s))$ and we want to construct a list of $\equiv$-inequivalent $1$-extensions belonging to a class $\mathscr P^q(i,(r,r),K,(s+1,s+1))$. The exact techniques used will depend heavily on the choice of $\equiv$. There are techniques in literature that allow the construction of distinct $1$-extensions linear extensions for $K=\{0,1\}$, i.e. the case of $\mbb F_q$-representable matroids.  These procedures use \textit{chains} of a matroid and are implemented in \textsf{sage-matroid} package of \textsf{SageMath}\cite{sage-matroid}. Isomorph  rejection via pairwise isomorphism testing is then employed to reject strongly isomorphic $1$-extensions. Note that approach using the chains of a matroid can also be extended to procedure $\textsf{nse}(\cdot)$. 

An alternative approach, which allows arbitrary sets $K$ of singleton ranks (and hence, to representably polymatroids as opposed to only representable matroids) is based on Leiterspiel, or the algorithm of snakes and ladders. This is a very general approach, designed for determining representatives of the orbits under action of a group $G$ in the power set $2^{\mathscr X}$ of a set $\mathscr X$. It can be used for generating any combinatorial objects that can be described as subsets of a set up to an equivalence relation that arises from the orbits of a group action on the set in question.  It was first described in a purely group-theoretic language by Schmalz \cite{schmalzleiter}, whereas an interpretation more suitable to combinatorial generation can be found in \cite{betten2006error}. Each member of $\mathscr P^q(i,(r,r),K,(s+1,s+1))$ can be described as an $i$-subset of $\tsf{Gr}_q(r,K)$.  Furthermore, weak isomorphism is in fact defined via a group action on $\tsf{Gr}_q(r,K)$, as seen in \S\ref{sec:polyiso}. To authors' knowledge, weak isomorphism is maximal amoungst all equivalence relations that can be defined via a group action on $\tsf{Gr}(q,K)$, in a terms of coarseness of the partition induced on  subsets of size $i$ of $\tsf{Gr}_q(r,K)$. In exchange for weakness of the isomorphism relation, Leiterspiel allows us avoid explicit pairwise isomorphism testing along with providing access to subgroups  of automorphism groups of polymatroids, which are constructed naturally in the process. These symmetries of polymatroids can be used in the determination of $p\mc I$-feasibility as described in \S\ref{sec:sympmap}. Owing to its generality, and other advantages, the implementation \tsf{ITAP} accompanying this article uses Leiterspiel for the procedure $\tsf{se}(\cdot)$. 

Next, for implementing the procedure $\tsf{nse}$, one can first form all distinct $\vert \tsf{us}(P)\vert$ non-simple $1$-extensions of each polymatroid $P$ in the input list and resort to pairwise isomorphism testing with respect to the chosen equivalence relation.  Note that lemma \ref{nsiso}, dictates how explicit strong isomorphism testing for two polymatroids $P_1$ and $P_2$ with $\tsf{us}(P_1)=\tsf{us}(P_2)$ can be performed. The statement of lemma \ref{nsiso} can be modified when one is interested in weak isomorphism testing, by substituting words 'weak' for 'strong' and by restring the meaning of automorphism group to be a subgroup of appropriate projective semilinear group.  On the flipside, if we know beforehand that  $\tsf{us}(P_1)\neq\tsf{us}(P_2)$, we can directly conclude that $P_1\not\cong P_2$ (alternatively, $P_1\not\overset{W}{=}P_2$). This allows us to restrict the explicit isomorphism testing to only the polymatroids with the same underlying simple polymatroid. Note that if Leiterspiel is used to construct simple polymatroids (as is done in \tsf{ITAP}), we have access to the automorphism group w.r.t. weak isomorphism of every simple polymatroid we construct. This gives us a head start in the application of lemma \ref{nsiso} as we are already aware of the automorphism group of $\tsf{us}(P_1)$ ($=\tsf{us}(P_2))$ w.r.t weak isomorphism, while we have a subgroup of automorphism group w.r.t. strong isomorphism.

When we are solving CLRP$_q$-EN to compute achievable network coding rate regions, we can also use a hybrid approach of switching between isomorphism relations, which is implemented in \tsf{ITAP}. This approach  uses Leiterspiel to perform simple extensions (green arrows in the first horizontal level in Fig. \ref{clrp_codecon_strategy}) and uses explicit isomorph testing w.r.t strong ismorphism relation when performing non-simple extensions (blue arrows in Fig. \ref{clrp_codecon_strategy}). In this case the automorphism groups provided by Leiterspiel can be used in explicit strong isomorphism testing.
This hybrid technique ultimately answers CLRP$_q$-EN with the strong isomorphism (i.e. equality of polymatroid rank vectors) notion of equivalence, even though intermediate stages -- the simple extensions -- make use of weak isomorphism equivalence relations.  
  This approach suffices for computing the achievable rate regions, because all rate vectors achievable with codes in $\mathscr P^q(\mb c)$ can be obtained from the set of all strongly non-isomorphic $p\mc I$-polymatroids in $\mathscr P^q(\mb c)$. 

\begin{algorithm}[h!]
\caption{An algorithm to solve CLRP$_q$-EN as implemented in \tsf{ITAP}}
\label{itapalgo}
\DontPrintSemicolon 
\KwIn{$\mc I$, a prime power $q$, $\mb c=(N,(r_l,r_u),K,(s_l,s_u))$ and vector space dimension $r$}
\KwOut{Lists $\mathfrak P_{N,s},s_l\leq s\leq s_u$ of codes in $\mathscr P^q(\mb c)$ and respective certificate maps $\tsf{cert}_{N,s}:\mathfrak P_{N,s}\rightarrow T_N$}
$\mathfrak P_{0,0}\leftarrow \{P_{\tsf{null}}\}$\\
$\tsf{cert}_{0,0}(P_{\tsf{null}})\leftarrow \phi_{\tsf{null}}$\\
\For{$1\leq i\leq s_u$}
{
	$(\mathfrak P_{i,i},\sigma_{i,i},\varphi_{i,i})\leftarrow \text{leiterspiel}(\mathfrak P_{i-1,i-1},\sigma_{i-1,i-1},\varphi_{i-1,i-1})$\;\label{ln:leiter}
	$(\mathfrak P_{i,i},\tsf{cert}_{i,i})\leftarrow \text{pmapfilter}( P_{i,i},\tsf{cert}_{i-1,i-1},\sigma_{i-1,i-1})$\;\label{ln:filter1}
	\If{$i\geq s_l+1$}
	{
		\For{$s_l\leq j \leq i-1$}
		{
			$\mathfrak P_{i,j}\leftarrow \text{nse}(\mathfrak P_{i-1,j})$\;\label{ln:nse1}
			$(\mathfrak P_{i,j},\tsf{cert}_{i,j})\leftarrow \text{pmapfilter}( P_{i,j},\tsf{cert}_{i-1,j},\sigma_{j,j})$\;\label{ln:filter2}
		}
		
	}
	\For{$s_u+1\leq i\leq N$}
	{
		\For{$s_l\leq j \leq s_u$}
		{
			$\mathfrak P_{i,j}\leftarrow  \text{nse}(\mathfrak P_{i-1,j})$\;\label{ln:nse2}
			$(\mathfrak P_{i,j},\tsf{cert}_{i,j})\leftarrow \text{pmapfilter}( P_{i,j},\tsf{cert}_{i-1,j},\sigma_{j,j})$\;\label{ln:filter3}
		}
	}
	
}
\Return{ $\{\mathfrak P_{N,s_l},\hdots,\mathfrak P_{N,s_u}\},\{\tsf{cert}_{N,s_l},\hdots,\tsf{cert}_{N,s_u}\}$}
\end{algorithm}

Algorithm \ref{itapalgo} provides a detailed description of how to implement the general strategy in fig. \ref{clrp_codecon_strategy}. It closely matches our implementation in \tsf{ITAP}. We assume that we are given a collection of constraints $\mc I$, size of finite field $q$, and a class tuple $\mb c=(N,(r_l,r_u),K,(s_l,s_u))$. Algorithm \ref{itapalgo} describes the construction for a specific $r_l\leq r\leq r_u$. The output is the a list of all weakly non-isomorphic $\mc I$-polymatroids in $\mathscr P^q(N,(r,r),K,(s_l,s_u))$. A list of weakly non-isomorphic polymatroids is denoted as $\mathfrak P_{i,j}$, where $i$ is the size of polymatroids in the list and $j$ is the size of underlying simple polymatroid of every polymatroid in the list. At the $i$th iteration of the algorithm, a collection of such lists of polymatroids of size $i$ are created from a collection of lists of polymatroids of size $i-1$. For every list $\mathfrak P_{i,j}$, a certificate map $\tsf{cert}_{i,j}$ is also maintained, which maps members of $\mathfrak P_{i,j}$ to the vertices of the $p$-map tree $T_i$ at depth $i$, that correspond to the certificates of $p\mc I$-feasibility.  The procedure leiterspiel($\cdot$) (line \ref{ln:leiter}) refers to the Leiterspiel algorithm as described in \cite{betten2006error} (Algorithm 9.6.10), which serves as a concrete implementation of procedure se($\cdot$), which we mentioned previously, without giving any internal details (green arrows in fig. \ref{clrp_codecon_strategy}). The input to this procedure is the orbits datastructure, consisting of a list $\mathfrak P_{i,i}$ for some $1\leq i\leq N$, stabilizer map $\sigma_{i,i}$ and the transporter map $\varphi_{i,i}$. The stabilizer map  $\sigma_{i,i}$ maps the members of $\mathfrak P_{i,i}$ to subgroups of $P\Gamma L(r,q)$ that are their automorphism groups. The transporter map $\varphi_{i,i}$ maps a subset of size $i$ of $\tsf{Gr}_q(r,K)$ that is a $p\mc I$-polymatroid to the repesentative of its weak isomorphism class present in the list $\mathfrak P_{i,i}$. The procedure pmapfilter($\cdot$) in lines \ref{ln:filter1},\ref{ln:filter2} and \ref{ln:filter3} corresponds to the red boxes in fig \ref{clrp_codecon_strategy}. The input to this procedure is a list of polymatroids $\mathfrak P_{i,j}$, the associated certificate map $\tsf{cert}_{i,j}$ and the stabilizer map $\sigma_{j,j}$ which maps the underlying simple polymatroids of members of $\mathfrak P_{i,j}$ to the respective stabilizers. It uses extend\_pmap($\cdot$) (line \ref{ln:extendpmap}) to extend the certificate $p$-map of the parent polymatroid obtained using function $\tsf{parent}(\cdot)$, and rejects any polymatroids for which no such extension exists. Note that extend\_pmap($\cdot$) also takes the stabilizer subgroup of the underlying simple polymatroid of the polymatroid being tested, in order to exploit symmetry as described in \S\ref{sec:sympmap}.  Note that Leiterspiel, as described in \cite{betten2006error}, allows one to reject some of the generated objects if they do not satisfy an \textit{inherited test function}, which is an indicator function for a any inherited property of the objects being generated. Hence, in actual implementation, rejection of poymatroids that are not $p\mc I$-feasible in line \ref{ln:filter1} is performed naturally in procedure leiterspiel$(\cdot)$ itself. The procedure nse($\cdot$) (lines \ref{ln:nse1} and \ref{ln:nse2}) takes as input  a list of polymatroids and outputs all strongly non-isomorphic non-simple extensions of the polymatroids in the list. For each polymatroid in the input list, it constructs all non-simple extensions (line \ref{ln:allnsext} of \ref{proc:nse}) and rejects isomorphs using explicit strong isomorphism testing for polymatroids with identical underlying simple polymatroid (line \ref{ln:strongtest} of \ref{proc:nse}). 
\begin{procedure} [h]
\DontPrintSemicolon
\caption{pmapfilter($\mathfrak P,\tsf{cert},\sigma$) }\label{proc:pmapfilter}
	$\mathfrak P'\leftarrow \emptyset$\;
	\For{$P \in \mathfrak  P$}
			{
				$\phi\leftarrow \text{extend}\_\text{pmap}(P,\tsf{cert}(\tsf{parent}(P)),\sigma(\tsf{us}(P)))$\;\label{ln:extendpmap}
				\If{$\phi\neq\phi_{\tsf{null}}$}
				{
					$\mathfrak{P}'\leftarrow \mathfrak{P}'\cup \{P\}$\;
					$\tsf{cert}'(P)\leftarrow \phi$\;
				}
			}
	\Return{$\mathfrak P',\tsf{cert}'$}
\end{procedure}

\begin{procedure} [h]
\DontPrintSemicolon
\caption{nse($\mathfrak P$) }\label{proc:nse}
	$\mathfrak P'\leftarrow \emptyset$\;
	\For{$P\in \mathfrak P'$}
	{
		$i\leftarrow$ size of $\tsf{us}(P)$\;
		$\mathfrak P''\leftarrow$ $ i+1$ non-simple extensions of $P$\;\label{ln:allnsext}
		\For{$P_{\tsf{ext}}\in \mathfrak P''$}
		{
			\tsf{badpoly} $\leftarrow \tsf{false}$\;
			\For{$P_{\tsf{ext}}'\in \mathfrak P'$}
			{
				\If{$\tsf{us}(P_{\tsf{ext}})= \tsf{us}(P_{\tsf{ext}}')$ $\land$ $P_{\tsf{ext}}\cong P_{\tsf{ext}}'$\label{ln:strongtest}}
				{
					\tsf{badpoly} $\leftarrow \tsf{true}$\; 
				}
			}
			\If{\tsf{badpoly} $= \tsf{false}$}
			{
				$\mathfrak P'\leftarrow \mathfrak P'\cup \{P\}$\;
			}
		}
	}
\Return{$\mathfrak P'$}
\end{procedure}

\section{Computational Experiments}\label{sec:example}
Accompanying this article is an implementation of the strategy described in Fig. \ref{clrp_codecon_strategy} written in GAP \cite{GAP4} that is available in form of a \tsf{GAP4} package named \tsf{ITAP} (Information Theoretic Achievability Prover) \cite{jayantitap}. It uses Leiterspiel \cite{betten2006error} to perform simple linear extensions. In this section we first consider a simple approach to measure the difficulty of solving a variant of CLRP$_q$-EN for a specific collection of constraints $\mc I$. We also consider several examples from literature to describe the functionality of \tsf{ITAP} via sample sessions in \tsf{ITAP}.

We will assume that every polymatroid generated using strategy in fig. \ref{clrp_codecon_strategy} is endowed with a \textit{rank oracle} i.e. a computer program that provides the rank of a subset of subspaces in time $\mc O(1)$. Denote by $\tsf{RE}_{\mc I}(i,r,q,K)$, the number of evaluations of the rank oracle performed while constructing $p\mc I$-polymatroids at iteration $i$ from $p\mc I$-polymatroids constructed at iteration $i-1$, for specific values of $r,q$ and $K$, for a specific collection of constraints $\mc I$. $\tsf{RE}_{\mc I}(i,r,q,K)$ can be written as,
\begin{equation}\label{eq:re_complexity}
\tsf{RE}_{\mc I}(i,r,q,K)=\tsf{RE}' _{\mc I}(i,r,q,K)\times \mc N_{i-1}^{r,q,K}, i\in[N]
\end{equation}
where, $\tsf{RE}_{\mc I}'(i,r,q,K)$ is the number of evaluations of the rank oracle per object and $\mc N_{i-1}^{r,q,K}$ denotes number of $p\mc I$-polymatroids at iteration $i$. Note that the first term in the expression above depends on the low level implementation and the collection of constraints $\mc I$. The second term, however, depends completely on the constraints $\mc I$. 
Hence, we are motivated to discuss the number of $p\mc I$-polymatroids constructed by \tsf{ITAP} as a measure of difficulty of solving a CLRP$_q$ variant for constraints $\mc I$. A closed form expression for the numbers of strongly non-isomorphic $p\mc I$-polymatroids in a particular class of codes for a particular collection of constraints $\mc I$ is unknown. Note that the numbers of $p\mc I$-polymatroids in a particular class of codes constructed by \tsf{ITAP} upper bounds the number of strongly non-isomorphic $p\mc I$-polymatroids in a particular class of codes, due to the hybrid approach of switching between weak and strong isomorphism adopted by \tsf{ITAP} (see \S \ref{sec:lowlevel}). Interestingly, we observe that these numbers are much smaller than the number of all codes up to strong isomorphism in the same class of codes, at least in cases where such upper bounds are known. 

\begin{figure}
\begin{center}
\includegraphics[angle=90,scale=0.16]{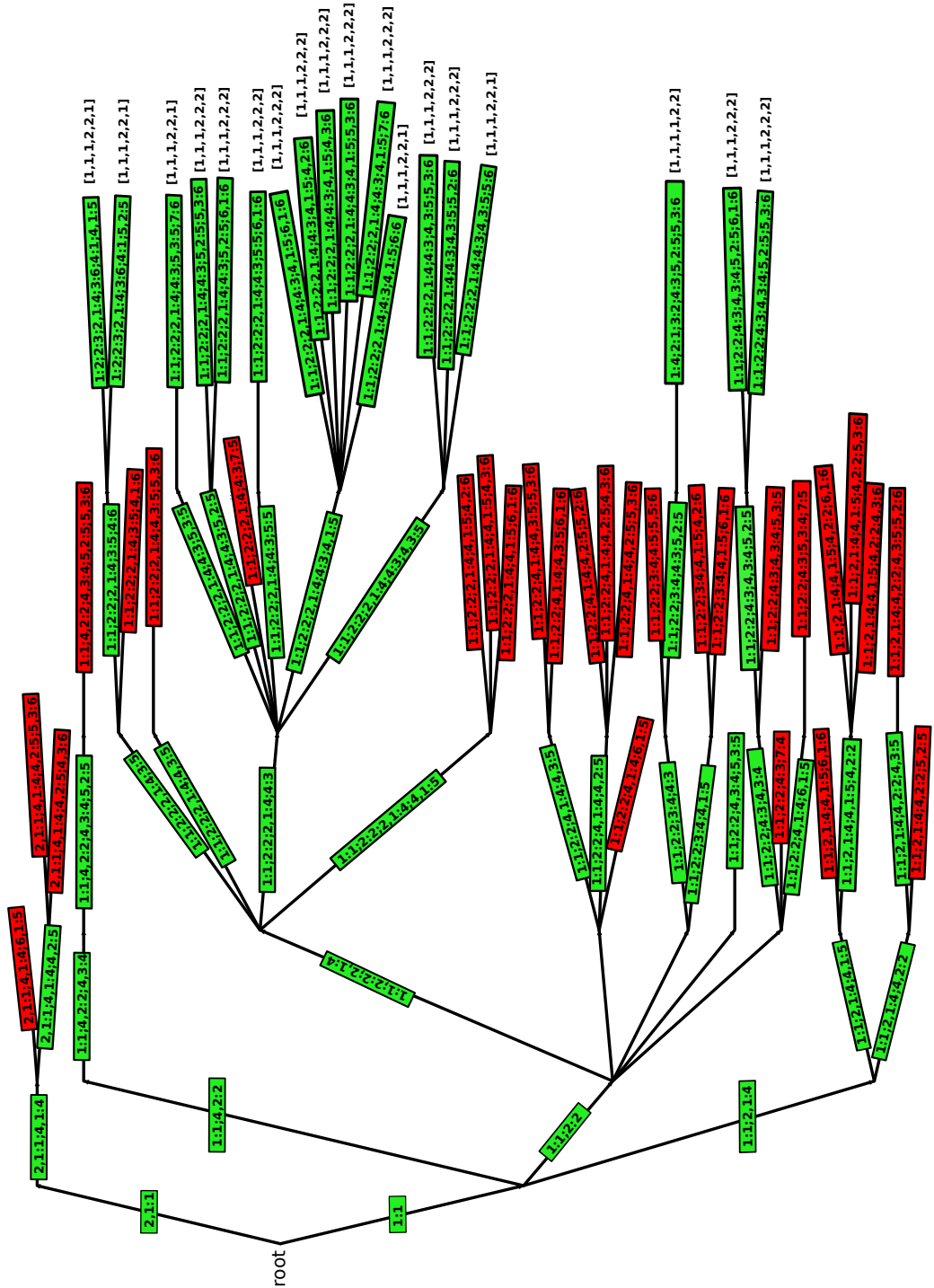}
\end{center}
\caption{The generation tree for example \ref{ex5} with class of codes $\mathscr P^2((6,(3,3),{1,2},(6,6)))$. The strings associated with edges encode the polymatroids and associated $p$-maps in a compact form. A string i,j:k;l,m:n  is to be interpreted as the polymatroid associated with the subspace arrangement \{\{i,j\},\{l,m\}\} where numbers i,j,l,m correspond to integer representation of binary vectors in $\mbb F_2^3$ and numbers $k,n$ correspond to subscripts of the random variables associated with the network.}
\label{gentree}
\end{figure}

The examples we consider, along with many others, are inbuilt in \tsf{ITAP} as part of the catalog of examples. The first example we consider is that of enumeration of all rate vectors achievable with a specified class of codes. The rest of the examples are concerned with the existential questions arising in varied contexts ranging from achievability proofs in network coding and secret sharing to proving linear rank inequalities. For each example, we state the associated constrained linear representability problem, plot the number of polymatroid representations constructed at the $i$th iteration of the algorithm along with known upper bounds, specify the time required to compute the answer, and describe the associated sample session with \tsf{ITAP}. All computations are performed on a 2 GHz Xeon CPU E5-2620 running Ubuntu 12.04 OS.
\begin{figure}[h!]
\begin{center}
\includegraphics[scale=1.1]{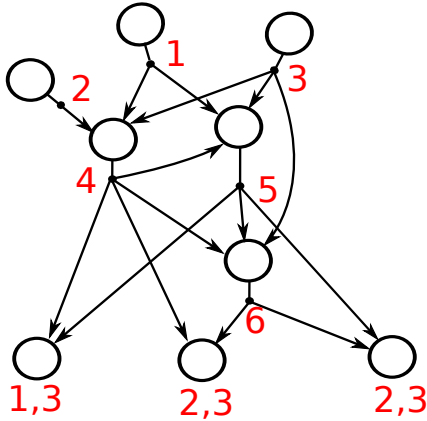}
\end{center}
\caption{A HMSNC instance $\tsf{HN1}$ with 6 random variables considered in example \ref{ex5} }
\label{rrexample}
\end{figure}
\begin{exmp}\label{ex5}
Consider the 6-variable HMSNC instance \tsf{HN1} in Fig. \ref{rrexample}. It consists of 3 source random variables and 3 edge random variables. 
The exact rate region of this network is also shown in below, which happens to be polyhedral.
\begin{equation} \label{rrex}
\mc R_{\tsf{HN1}}=\left\{(\bs \omega, \mb r)\in \mbb R^6\left|\begin{aligned}
R_4\geq \omega_1\\
R_4+R_5\geq \omega_1+\omega_2+\omega_3\\
R_6\geq \omega_1\\
R_4+R_6\geq h_2+\omega_3\\
R_4+R_5+2R_6\geq  \omega_1+2\omega_2+2\omega_3\\
R_5+R_6\geq \omega_2+\omega_3
\end{aligned}\right.\right\}
\end{equation}
 The network constraints associated with this network are,
\begin{equation} \label{hn1con}
\mathcal I_{\tsf{HN1}}=\left\{\begin{aligned}
\{h_1+h_2+h_3=h_{1,2,3}\},\{h_{ 1, 2, 3  } = h_{ 1, 2,3, 4 }\},\\ \{h_{ 1, 3,4 } = h_{ 1, 3, 4,5 }\},
 \{h_{ 3,4, 5 } = h_{ 3,4, 5, 6 }\},\\ \{h_{ 4, 5 } = h_{ 1,3, 4, 5 }\},
 \{h_{ 4, 6 } = h_{ 2,3, 4, 6 }\},\\  \{h_{ 5,6 } = h_{ 2,3,5, 6 }\}
\end{aligned}\right\}
\end{equation}
We can construct achievable rate region that matches the exact rate region in \eqref{rrex} if we use codes from class $\mathscr P^q(6,(1,4),\{0,1,2\},(1,6))$, as shown in the sample \tsf{ITAP} session. This computation takes about 387 sec. Constructing an achievable rate region from class $\mathscr P^q(6,(1,3),\{0,1,2\},(1,6))$, however, results in a smaller rate region, shown in \eqref{rrin}, computation that takes about 32 sec. The tree of $p\mc I_{\tsf{HN1}}$-polymatroids generated by \tsf{ITAP} in the latter case is shown in Fig. \ref{gentree}. The leaves of this tree correspond to all weakly non-isomorphic network codes that belong to the family of representable integer polymatroids determined by the aforementioned parameters. Each leaf comes with a $p$-map that determines the rate vector achieved by the associated code, which is also shown in the figure. This $p$-map has the property that it is the lexicographically smallest map amoung all valid $p$-maps. By traversing rest of the $p$-map tree for these $\mc I_{\tsf{HN1}}$-polymatroids, we recover the following collection of achievable rate vectors:
\begin{equation}
\left\{\begin{aligned} [ 1, 1, 1, 1, 2, 2 ],[ 1, 1, 1, 2, 1, 2 ],\\
 [ 1, 1, 1, 2, 2, 1 ],   [ 1, 1, 1, 2, 2, 2 ]\end{aligned}\right\}
\end{equation}  
To obtain the inequality description of the achievable rate region, we follow the procedure mentioned at the end of \S\ref{sec:clrpnc}. This completes the computation of achievable rate region.

\begin{equation} \label{rrin}
\mc R_{\tsf{in}}=\left\{(\bs \omega, \mb r)\in \mbb R^6\left|\begin{aligned}
\omega_i \geq 0,i\in[3]\\
R_i\geq\omega_k,i\in\{4,5,6\},k\in[3] \\
 R_i  +R_j \geq 3\omega_k,\forall \{i,j\}\subset \{4,5,6\}\text{ and }k\in[3]  \\
 R_4  +R_5  +R_6\geq 5\omega_i,i\in[3] 
\end{aligned}\right.\right\}
\end{equation}
\begin{Verbatim}[commandchars=!|B,fontsize=\small,frame=single,label=ITAP session with example \ref{ex5}]
  !gapprompt|gap>B !gapinput|N:=HyperedgeNet1();B
  [ [ [ [ 1, 2, 3 ], [ 1, 2, 3, 4 ] ], [ [ 1, 3, 4 ], [ 1, 3, 4, 5 ] ],
        [ [ 3, 4, 5 ], [ 3, 4, 5, 6 ] ], [ [ 4, 5 ], [ 1, 3, 4, 5 ] ],
        [ [ 4, 6 ], [ 2, 3, 4, 6 ] ], [ [ 5, 6 ], [ 2, 3, 5, 6 ] ] ], 3, 6 ]
  !gapprompt|gap>B !gapinput|rlist:=proveregion(N,2,GF(2),[4]);; # k=2,opt_dmax=4=max. code dimensionB
  !gapprompt|gap>B !gapinput|Size(rlist[1]); # number of distinct achievable rate vectors foundB
  122
  !gapprompt|gap>B !gapinput|rlist[1][78]; # an achievable rate vectorB
  [ 2, 0, 1, 2, 1, 1 ]
  !gapprompt|gap>B !gapinput|lrs_path:="/home/aspitrg3-users/jayant/lrslib-061/";; # path to lrslibB
  !gapprompt|gap>B !gapinput|rrcompute(rlist[1],N[2],N[3],lrs_path); # compute achievable rate regionB
  
  *redund:lrslib v.6.1 2015.11.20(lrsgmp.h gmp v.5.0)
  *Copyright (C) 1995,2015, David Avis   avis@cs.mcgill.ca
  *Input taken from file /tmp/tmxYdXYT/file1.ext
  *Output sent to file /tmp/tmxYdXYT/file1red.ext
  
  *0.056u 0.004s 648Kb 0 flts 0 swaps 0 blks-in 8 blks-out
  
  
  *lrs:lrslib v.6.1 2015.11.20(lrsgmp.h gmp v.5.0)
  *Copyright (C) 1995,2015, David Avis   avis@cs.mcgill.ca
  *Input taken from file /tmp/tmxYdXYT/file1red.ext
  H-representation
  begin
  ***** 7 rational
   0  0  0  0  1  0  0
   0  1  0  0  0 -1  0
   0  0  0  0  0  1  0
   0  0  0  0  0  0  1
   0  0  0  1  0  0  0
   0  1  1  0 -1 -1 -1
   0  0  1  1  0 -1 -1
   0  0  1  0  0  0  0
   0  1  1  2 -1 -2 -2
   0  1  0  1  0 -1 -1
  end
  *Totals: facets=10 bases=22
  *Dictionary Cache: max size= 6 misses= 0/21   Tree Depth= 5
  *lrs:lrslib v.6.1 2015.11.20(32bit,lrsgmp.h)
  *0.000u 0.000s 648Kb 0 flts 0 swaps 0 blks-in 0 blks-out
\end{Verbatim}

\end{exmp}
\begin{exmp} \label{ex:matroidal}
This is a collection of examples from network coding literature: the so called matroidal and discrete polymatroidal networks (see \cite{DFZMatroidNetworks,bsrajan14}). These networks are constructed to mimic the dependencies of matroids and integer polymatroids respectively. As a result, the known results regarding the representability of these polymatroids carry forward to the networks, thus providing us with the networks for which achievability or non-achievability of certain rate vectors is established by construction. The matroidal networks we consider are the Fano, Non-Fano and V\'amos networks, whereas the discrete polymatroidal network we use as example is the network constructed from the polymatroid associated with a scaled version of $U^2_4$ matroid.  
Equations below describe the constraints associated with each of these networks. The Fano network is size $N=7$ network associated with the Fano matroid. Rate vector $(\omega_i=1,R_j=1,i\in [3], j\in[7]\setminus [3])$ is achievable for this network using linear network coding only over a finite field of even characteristic, as the Fano matroid is only representable over such fields.  
\begin{equation} \label{fanocon}
\mc I_{\tsf{Fano}}=\left\{\begin{aligned}
\{h_1+h_2+h_3=h_{1,2,3}\},\{h_{ 1, 2 } = h_{ 1, 2, 4 }\},\\ \{h_{ 2, 3 } = h_{ 2, 3, 5 }\},
 \{h_{ 4, 5 } = h_{ 4, 5, 6 }\},\\ \{h_{ 3, 4 } = h_{ 3, 4, 7 }\},
 \{h_{ 1, 6 } = h_{ 3, 1, 6 }\},\\  \{h_{ 6, 7 } = h_{ 2, 6, 7 }\}, \{h_{ 5, 7 } = h_{ 1, 5, 7 }\}
\end{aligned}\right\}
\end{equation}
\begin{Verbatim}[commandchars=!@|,fontsize=\small,frame=single,label=ITAP session with Fano Network in Example \ref{ex:matroidal}]
  !gapprompt@gap>| !gapinput@FanoNet();|
  [ [ [ [ 1, 2 ], [ 1, 2, 4 ] ], [ [ 2, 3 ], [ 2, 3, 5 ] ],
       [ [ 4, 5 ], [ 4, 5, 6 ] ], [ [ 3, 4 ], [ 3, 4, 7 ] ],
       [ [ 1, 6 ], [ 3, 1, 6 ] ], [ [ 6, 7 ], [ 2, 6, 7 ] ],
       [ [ 5, 7 ], [ 1, 5, 7 ] ] ], 3, 7 ]
  !gapprompt@gap>| !gapinput@rlist:=proverate(FanoNet(),[1,1,1,1,1,1,1],GF(2),[]);;|
  !gapprompt@gap>| !gapinput@rlist[1]; # Fano matroid is representable over GF(2)|
  true
  !gapprompt@gap>| !gapinput@DisplayCode(rlist[2]);|
  1->1
   . . 1
  =============================
  2->2
   . 1 .
  =============================
  3->4
   . 1 1
  =============================
  4->3
   1 . .
  =============================
  5->6
   1 . 1
  =============================
  6->5
   1 1 .
  =============================
  7->7
   1 1 1
  =============================
  !gapprompt@gap>| !gapinput@rlist:=proverate(FanoNet(),[1,1,1,1,1,1,1],GF(3),[]);;|
  !gapprompt@gap>| !gapinput@rlist[1]; # Fano matroid is not representable over GF(3)|
  false
\end{Verbatim}
 
The second matroidal network we consider is the Non-Fano network, which is also a size $N=7$ network, for which the rate vector $(\omega_i=1,R_j=1,i\in [3], j\in[7]\setminus [3])$ is achievable via linear network coding only over a finite field of odd characteristic.   
\begin{equation} \label{nonfanocon}
\mc I_{\tsf{NonFano}}=\left\{\begin{aligned}
\{\{h_1+h_2+h_3=h_{1,2,3}\},\\
\{h_{1,2,3}=h_{1,2,3,4}\},\{h_{1,2}=h_{1,2,5}\},\\
\{h_{1,3}=h_{1,3,6}\},\{h_{2,3}=h_{2,3,7}\},\\
\{h_{4,5}=h_{3,4,5}\},\{h_{4,6}=h_{2,4,6}\},\\
\{h_{4,7}=h_{1,4,7}\},\{h_{5,6,7}=h_{1,2,3,5,6,7}\}
\end{aligned}\right\}
\end{equation}

The third matroidal network we consider is the  V\'amos network of size $N=8$ for which the rate vector $(\omega_i=1,R_j=1,i\in [4], j\in[8]\setminus [4])$ is not achievable via linear coding over any finite field. 
\begin{equation} \label{vamoscon}
\mc I_{\tsf{V\'amos}}=\left\{\begin{aligned}
\{h_1+h_2+h_3+h_4=h_{1,2,3,4}\},\{h_{1,2,3,4}=h_{1,2,3,4,5}\},\\
\{h_{1,2,5}=h_{1,2,5,6}\},\{h_{2,3,6}=h_{2,3,6,7}\},\\
\{h_{3,4,7}=h_{3,4,7,8}\},\{h_{4,8}=h_{2,4,8}\},\\
\{h_{2,3,4,8}=h_{1,2,3,4,8}\},\{h_{1,4,5,8}=h_{1,2,3,4,5,8}\},\\
\{h_{1,2,3,7}=h_{1,2,3,4,7}\},\{h_{1,5,7}=h_{1,3,5,7}\}
\end{aligned}\right\}
\end{equation}
The last network we consider is the  $2U^2_4$ network of size $n=4$, whose network constraints mimic the dependencies of the $U_4^2$ matroid i.e. the $4$ point line. This matroid is a forbidden minor for matroid representability over $\mbb F_2$\cite{OxleyMatroidBook}, which means that the rate vector $(\omega_i=1,R_j=1,i\in [2], j\in[4]\setminus [2])$ is not achievable for this network via linear network coding over $\mbb F_2$. However, the polymatroid $2U_4^2$, which 
is obtained by scaling the rank function of $U_4^2$ by 2, is linearly representable over $\mbb F_2$, implying that rate vector $(\omega_i=2,R_j=2,i\in [2], j\in[4]\setminus [2])$ is achievable for this network using linear coding over $\mbb F_2$.
\begin{equation} \label{u24con}
\mc I_{2U^2_4}=\left\{\begin{aligned}
\{h_1+h_2=h_{1,2}\},\{h_{ 1, 2 }=h_{ 1, 2, 3 }\},\\
 \{h_{ 1, 3 }=h_{ 1, 2, 3 }\}, \{h_{ 2, 3 }=h_{ 1, 2, 3 }\},\\
 \{h_{ 1, 2 }=h_{ 1, 2, 4 }\}, \{h_{ 1, 4 }=h_{ 1, 2, 4 }\},\\
  \{h_{ 3, 4 }=h_{ 1, 3, 4 }\}, \{h_{ 3, 4 }=h_{ 2, 3, 4 }\}, \{h_{ 2, 4 }=h_{ 1, 2, 4 }\}
\end{aligned}\right\}
\end{equation}

Now that we have described the four network coding instances, we can ask \tsf{ITAP} some questions
whose answers we already know. 
\begin{figure}[h!]

\begin{center}
\includegraphics[scale=0.6]{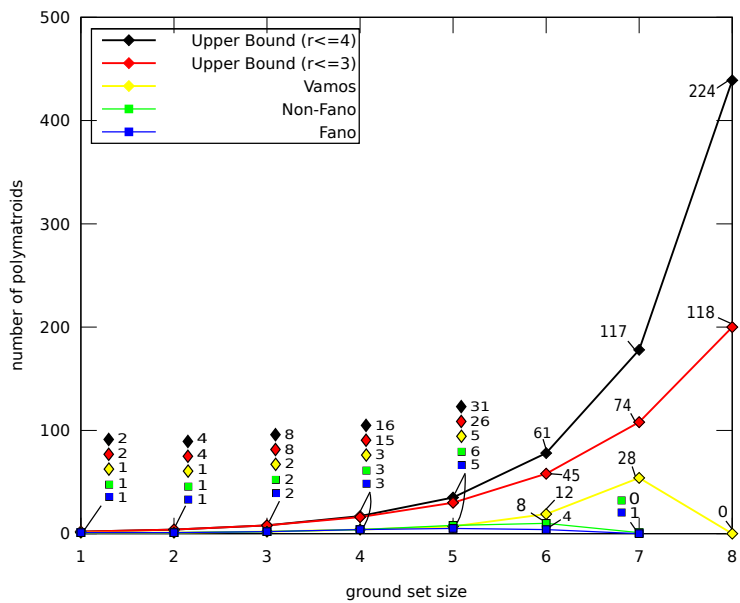}
\end{center}
\caption{Number matroid representations over $\mbb F_2$ maintained by \tsf{ITAP} at different iterations for Fano (blue), Non-Fano (green) and V\'amos (yellow) networks along with upper bound which is the number of all non-isomorphic $\mbb F_2$-representable matroids of rank $\leq 3$ for Fano and Non-Fano networks while it is the number of $\mbb F_2$-representable matroids of rank $\leq 4$ for the V\'amos network}
\label{f2_plot}
\end{figure}

\begin{figure}[h!]

\begin{center}
\includegraphics[scale=0.6]{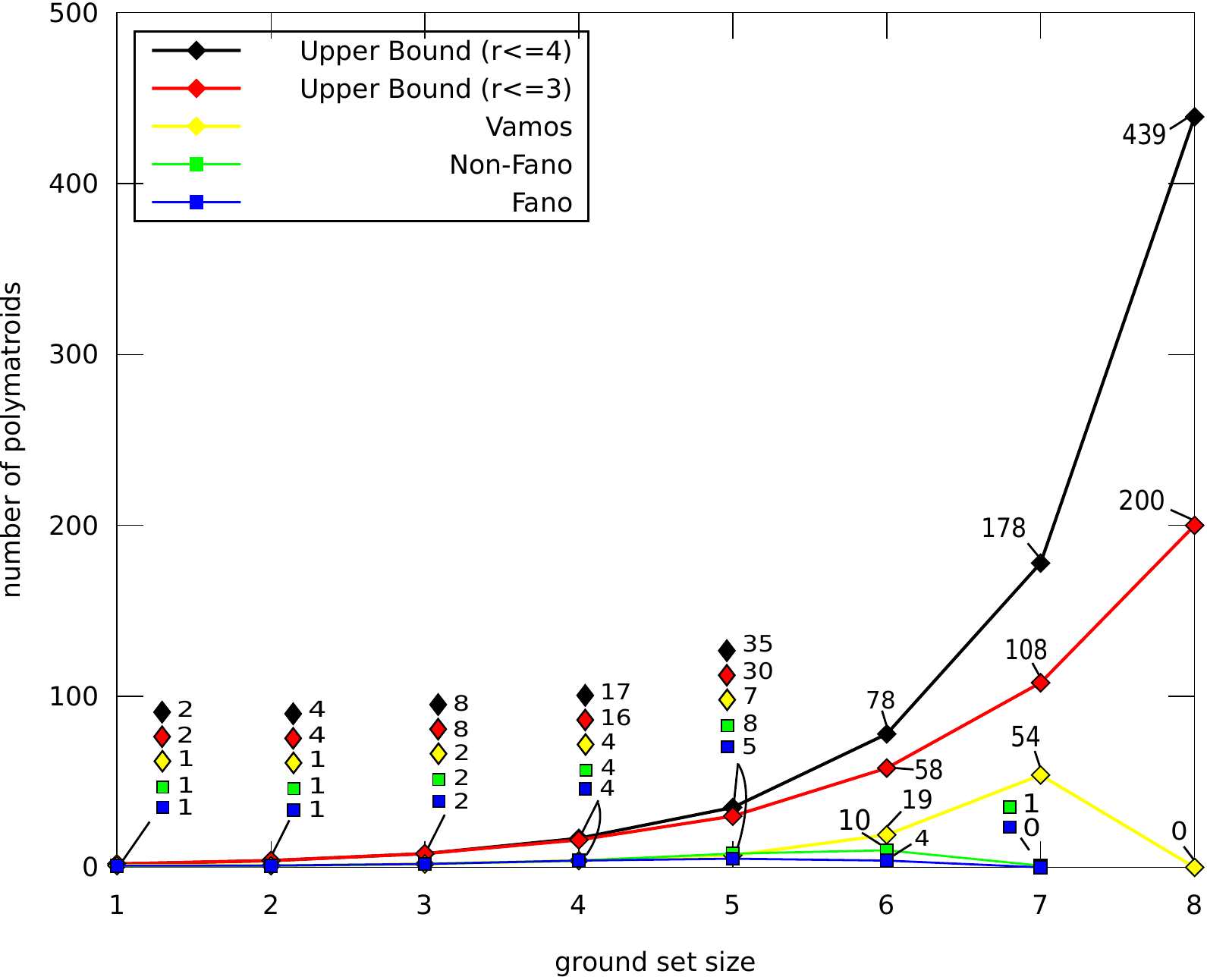}
\end{center}
\caption{Number of weakly non-isomorphic matroid representations over $\mbb F_3$ at different iterations for Fano, Non-Fano and V\'amos networks  along with upper bound (the number of all non-isomorphic $\mbb F_3$-representable matroids of suitable rank)}
\label{f3_plot}
\end{figure}

For Fano, Non-Fano and  V\'amos networks, we test whether rate vector $(\omega_i=1,R_j=1,i\in [k], j\in[N]\setminus [k])$ is achievable over $\mbb F_2$ and $\mbb F_3$. The sample session with for Fano network with \tsf{ITAP} is also shown. The numbers of $p\mc I$-polymatroids  are shown in figures \ref{f2_plot} and \ref{f3_plot}.  The upper bounds in these figures are the numbers of rank $\leq 3$ and rank $\leq 4$ non-isomorphic binary and ternary  representable matroids respectively, which were first obtained by  Wild (see \cite{wildenum,oeis_allbin}). One can see from the plots that the number $p\mc I$-polymatroids maintained by \tsf{ITAP} in each instance is much smaller than the respective upper bounds. The time required for testing scalar solvability of Fano and Non-Fano networks is 1 seconds and 2 seconds respectively over $\mbb F_2$ whereas it is 5 seconds and 4 seconds respectively over $\mbb F_3$. Timing results for V\'amos network are discussed in detail in example \ref{ex:grobcpmparison}.
 \begin{figure}
 \begin{center}
 \includegraphics[scale=0.6]{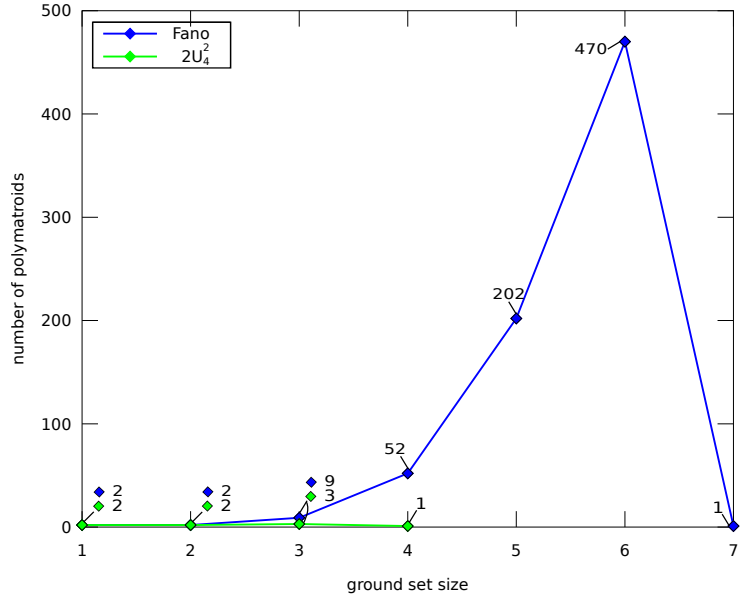}
 \end{center}
 \caption{Number of polymatroid representations over $\mbb F_2$ maintained by \tsf{ITAP} at different iterations for Fano and $2U^2_4$ networks}
 \label{k2f2_plot}
 \end{figure}
 
 For Fano and $2U_4^2$ networks, we test if rate vector $(\omega_i=2,R_j=2,i\in [k], j\in[N]\setminus [k])$ is achievable. We know that the answer is affirmative for both of these instances.  This rate vector dictates that for Fano network, the class of codes to search for achievability construction is $\mathscr P^q(7,(6,6),\{2\},(3,7))$. Whereas, for $2U_4^2$ network, we have the class $\mathscr P^q(4,(4,4),\{2\},(2,4))$. The result, i.e. the numbers of weakly non-isomorphic representations constructed at each iteration by \tsf{ITAP} is shown in figure \ref{k2f2_plot}.  This plot is left without any upper bounds, as the only known upper bound is the number of general $2$-polymatroids obtained via Savitsky's enumeration \cite{savitsky14}, already shown in Fig. \ref{polygrowth_oeis}. The time required in this case is 142 minutes and 1 seconds for Fano and $2U^2_4$ networks respectively.
$\blacksquare$
\end{exmp}
\begin{exmp}\label{ex_benaloh}(\hspace{-0.2mm}\cite{Benaloh90,padronotes})
Let $\Gamma=\{\{2,3\},\{3,4\},\{4,5\}\}$ be an access structure for sharing a secret with $4$ parties labeled $\{2,3,4,5\}$ with the dealer labeled $1$. This example appears in \cite{Benaloh90}, whereas an explicit multi-linear secret sharing scheme for this access structure with secret size $r_1=2$ and share sizes $r_2=2,r_3=3,r_4=3,r_5=2$ can be found in \cite{padronotes}. For the purpose of reproducing this scheme with \tsf{ITAP}, equations \eqref{sscon1} and \eqref{sscon2} (constraints $\mc I_\Gamma$) along with the requirement to have specified share sizes form collection of constraints $\mc I$. The constraints imply that the class of codes in which an achievable construction might exist is $\mathscr P^q(5,(3,12),\{2,3\},(2,5))$. Equation \eqref{benalohmatrix} gives the representation constructed by \tsf{ITAP}, with $p$-map $\{1\mapsto 1,2\mapsto 2,3\mapsto 3,4\mapsto 5,5\mapsto 4\}$. The sample \textsf{ITAP} session is also shown. This computation takes about 32 min. $\blacksquare$
\begin{equation}\label{benalohmatrix}
P\triangleq\left\{
\stackrel{\mathlarger{s_1}}{
\begin{bmatrix}
1 & 0\\
0 & 1\\
0 & 0\\
0 & 0\\
0 & 0\\
0 & 0
\end{bmatrix}},
\stackrel{\mathlarger{s_2}}{
\begin{bmatrix}
0 & 0\\
0 & 0\\
1 & 0\\
0 & 1\\
0 & 0\\
0 & 0
\end{bmatrix}},
\stackrel{\mathlarger{s_3}}{
\begin{bmatrix}
0 & 1 & 0\\
1 & 0 & 0\\
1 & 0 & 0\\
0 & 1 & 0\\
0 & 0 & 1\\
0 & 0 & 0
\end{bmatrix}},
\stackrel{\mathlarger{s_4}}{
\begin{bmatrix}
0 & 0\\
0 & 0\\
0 & 0\\
0 & 0\\
1 & 0\\
0 & 1
\end{bmatrix}},
\stackrel{\mathlarger{s_5}}{
\begin{bmatrix}
0 & 0 & 1\\
0 & 1 & 0\\
0 & 0 & 0\\
1 & 0 & 0\\
0 & 1 & 0\\
0 & 0 & 1
\end{bmatrix}}
\right\}
\end{equation}
\begin{Verbatim}[commandchars=!@|,fontsize=\small,frame=single,label=ITAP session for Example \ref{ex_benaloh}]
  !gapprompt@gap>| !gapinput@B:=BenalohLeichter();|
  [ [ 1, 2 ], [ 2, 3 ], [ 3, 4 ] ]
  !gapprompt@gap>| !gapinput@rlist:=provess(B,5,[2,2,3,3,2],GF(2),[]);;|
  !gapprompt@gap>| !gapinput@rlist[1];|
  true
  !gapprompt@gap>| !gapinput@DisplayCode(rlist[2]);|
  1->1
   . . . . 1 .
  . . . . . 1
  =============================
  2->2
  . . 1 . . .
  . . . 1 . .
  =============================
  3->3
  . 1 . . . .
  . . 1 . . 1
  . . . 1 1 .
  =============================
  4->5
  1 . . . . .
  . 1 . . . .
  =============================
  5->4
  1 . . . . 1
  . 1 . . 1 .
  . . 1 . . .
  =============================
\end{Verbatim}
\begin{figure}[h!]
\begin{center}
\includegraphics[scale=0.5]{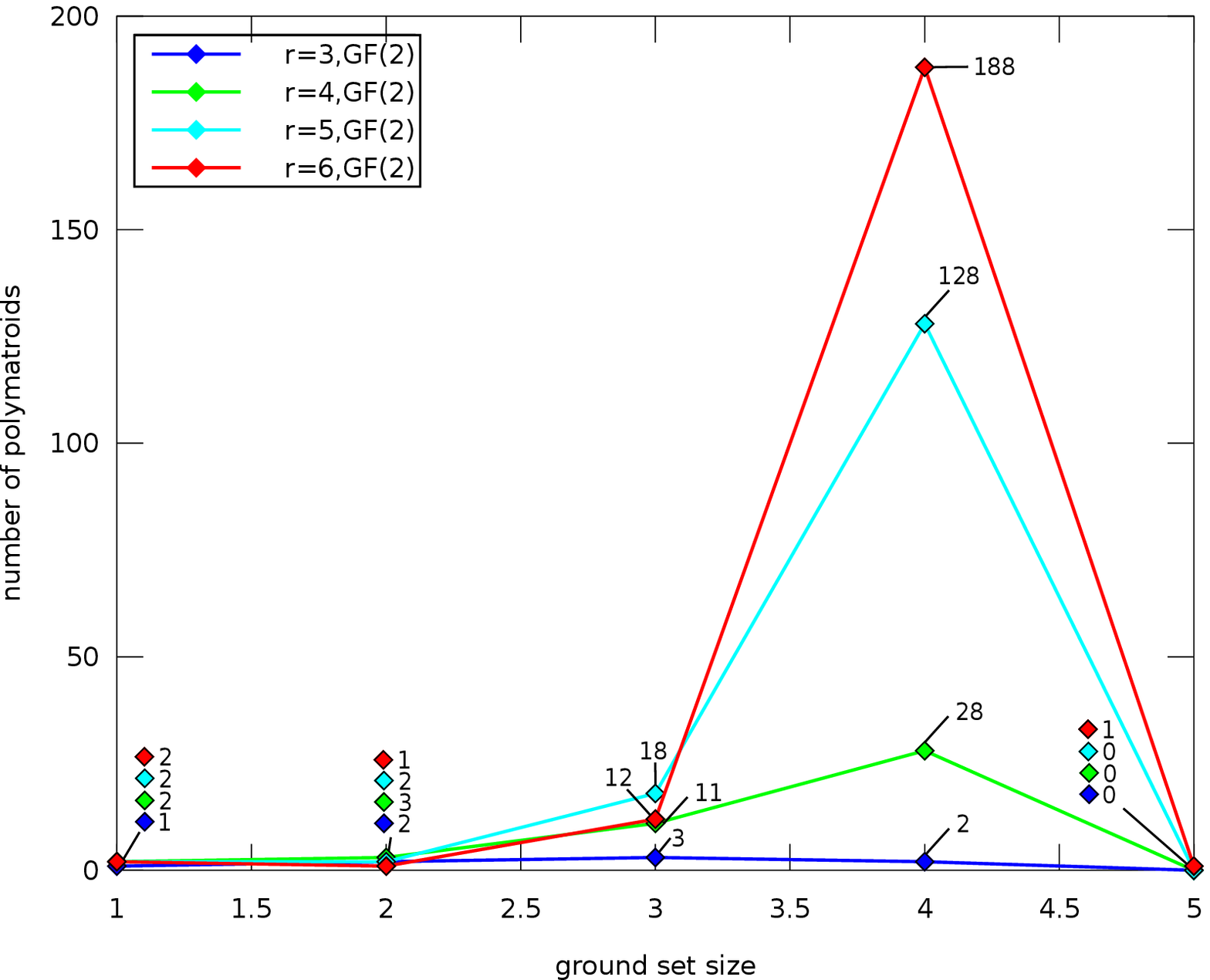}
\end{center}
\caption{Number of weakly non-isomorphic $p\mc I$-polymatroid representations over $\mbb F_2$ of various ranks constructed by \tsf{ITAP} at different iterations for example \ref{ex_benaloh}. }
\label{fanoex2_plot}
\end{figure}
\end{exmp}
\begin{exmp}\label{linrank6ex} This example is concerned with proving linear rank inequalities amoungst $N$ random variables. Dougherty, Freiling and Zeger use a computational technique \cite{dfz_linrank6} to find non-redundant linear rank inequalities in $N$ random variables (linear rank inequalities that cannot be expressed as conic combinations of the others). The idea is to start with the Shannon outer bound $\Gamma_{N'}$ for $N'>N$, intersect it with equality constraints obtained by enforcing certain pairs of random variables to have common information equal to the mutual information between them (currently, at most 3 pairs have been used for $N=6$ and $N'=7$) and then projecting down onto the dimensions associated with subsets of $N$ variables. The extreme rays of this cone that are linearly representable are also the extreme rays of the cone of the linear rank inequalities for $N$ random variables. This is where a computer based representability checker comes into picture (i.e. a computer program capable of solving problem (E2) described in \S\ref{sec:prelim}). A computer program like \tsf{ITAP}, in principle, is able to prove linear representability of such an extreme ray over $\mbb F_q$ by solving problem (E2$_q$) described in \S\ref{sec:prelim}. Consider the polymatroid $([6],f)$ where $f$ is specified by the vector $[1,1,2,1,2,2,2,2,3,3,4,3,4,4,4,2,3,3,4,3,4,4,4,4,5,5,6,5,6,6,6,4,5,5,6,5,6,6,6,6,6,6,6,\\
6,6,6,6,6,6,6,6,6,6,6,6,6,6,6,6,6,6,6,6]$, which is one of the extreme rays of the cone of linear rank inequalities \cite{dfz_linrank6} for $n=6$. 
The representation constructed by \tsf{ITAP} is shown in equation \eqref{linrank6mat} (with $p$-map $ \{\{i\mapsto i\},\spc\forall i\in [6]\}$).
\begin{equation}\label{linrank6mat}
P\triangleq\left\{
\stackrel{\mathlarger{s_1}}{
\begin{bmatrix}
1 \\
0 \\
0 \\
0 \\
0 \\
0
\end{bmatrix}},
\stackrel{\mathlarger{s_2}}{
\begin{bmatrix}
0 \\
1 \\
0 \\
0 \\
0 \\
0
\end{bmatrix}},
\stackrel{\mathlarger{s_3}}{
\begin{bmatrix}
1 \\
1 \\
0 \\
0 \\
0 \\
0
\end{bmatrix}},
\stackrel{\mathlarger{s_4}}{
\begin{bmatrix}
0 & 0 \\
0 & 0 \\
1 & 0 \\
0 & 1 \\
0 & 0 \\
0 & 0
\end{bmatrix}},
\stackrel{\mathlarger{s_5}}{
\begin{bmatrix}
0 & 0 \\
0 & 0 \\
0 & 0 \\
0 & 0 \\
1 & 0 \\
0 & 1
\end{bmatrix}},
\stackrel{\mathlarger{s_6}}{
\begin{bmatrix}
1 & 0 & 1 & 0 \\
0 & 1 & 0 & 1\\
0 & 0 & 0 & 1\\
0 & 0 & 1 & 0\\
0 & 1 & 0 & 0\\
1 & 0 & 0 & 0
\end{bmatrix}}
\right\}
\end{equation}
\begin{figure}[h!]

\begin{center}
\includegraphics[scale=0.5]{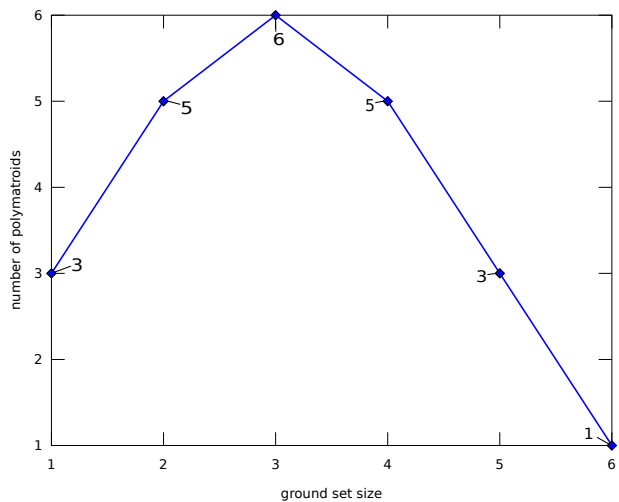}
\end{center}
\caption{Number of  $p\mc I$-polymatroid representations over $\mbb F_2$ of various ranks constructed by \tsf{ITAP} at different iterations for example \ref{linrank6ex}.}
\label{linrank6_plot}
\end{figure}
Figure \ref{linrank6_plot} shows the number of weakly non-isomorphic representations of $p\mc I$-polymatroids constructed by 
\tsf{ITAP} at each iteration while solving (E2$_q$) w.r.t. the aforementioned ray. This computation takes about 375 sec. A similar strategy has been used by authors in \cite{apteitw16} to prove tightness of symmetric parts of $\Gamma_N$ on symmetric parts of $\overline{\Gamma_N^*}$, \tsf{ITAP} was used to prove entropicness of the extreme rays of symmetric parts of $\Gamma_N$ under the action of groups permuting the random variables. 
 $\blacksquare$

\end{exmp}
\begin{exmp}\label{ex:grobcpmparison}
In this example, we compare the computational performance of methods developed in this work and the Gr\"obner basis compution basis based methods, in particular the path gain construction of Subramanian and Thangaraj\cite{subrapathgain10}, for solving \tsf{CLRP}$_q$-EX. This method is implemented in \tsf{ITAP} via the \tsf{GAP} interface to singular \cite{singulargap} which is available as an add on package to \tsf{GAP}, and interfaces to popular Gr\"obner basis computation software \tsf{Singular}\cite{DGPS}. The algorithm that transforms a given HMSNC instance to an instance of Network Coding over Direct Acyclic Multi-graphs (NCDAMG) is provided in appendix \ref{app:transform}. The first instance for which we compare the two methods is the V\'amos network.Being a 'no' instance of \tsf{CLRP}$_q$-EX, this is expected to trigger the worst behaviour out of a combinatorial generation based algorithm (in a sense that the algorithm must traverse the entire search tree, in order to conclude that a solution does not exist). Our results are summarized in table X, which suggest the same. 
\begin{table}
\begin{center}
\begin{tabular}{|c|c|c|}
\hline
q & Comb. Gen. & Gr\"obner\\
\hline
2 &  130 &  4\\
2 &  1330  & 5\\
4 &  5400 &  7\\
5 &  >2 hrs &  5\\
7 &  >2 hrs &  5\\
8 &  >2 hrs &  7\\
9 &  >2 hrs &  7\\
\hline
\end{tabular}
\caption{Time in seconds to test scalar solvability of V\'amos network w.r.t. Field size $q$}
\end{center}
\end{table}
One the other hand, we also find type of instances that incite bad behaviour from Gr\"obner basis computation based method. The number of indeterminates in path gain formulation depends on the number of paths. If the NCDAMG instance produced by algorithm \ref{transformhmsnc} has $p$ source-sink paths for a rate vector $(\omega_i=1,R_j=1),i\in[k],j\in[N]\setminus [k]$, doubling each rate produces a NCDAMG instance with $2^p$ paths. The instance we compare the performance of the two methods is a Multilevel Diversity Coding System (MDCS) \cite{YeungBook} instance with $N=7$, described as follows:
\begin{equation} \label{mdcscon}
\mc I_{\tsf{MDCS}}=\left\{\begin{aligned}
\{\{h_1+h_2+h_3=h_{1,2,3}\},
\{h_{1,2,3}=h_{1,2,3,4}\},\\
\{h_{1,2,3}=h_{1,2,3,5}\},
\{h_{1,2,3}=h_{1,2,3,6}\},\\
\{h_{1,2,3}=h_{1,2,3,7}\},
\{h_{4}=h_{1,4}\},
\{h_{5}=h_{1,5}\},\\
\{h_{4,5}=h_{1,2,4,5}\},
\{h_{6,7}=h_{1,2,6,7}\},\\
\{h_{4,6}=h_{1,2,3,4,6}\},
\{h_{5,7}=h_{1,2,3,5,7}\}
\end{aligned}\right\}
\end{equation}

\begin{table}
\begin{center}
\begin{tabular}{|c|c|c|c|}
\hline
Rate vector & Type of instance & Comb. Gen. & Gr\"obner\\
\hline
$[1,1,1,1,1,1,1]$ & no &  4 & 1\\
$[1,1,1,2,1,1,1]$ & no &  8  & 1\\
$[1,1,1,2,2,1,1]$& yes &  9 &  >1 hr\\
$[1,1,1,2,2,2,2]$ & yes & 3 & >1 hr\\
$[1,1,1,2,1,1,2]$ & yes & 5 & >1 hr\\
\hline
\end{tabular}
\caption{Time in seconds to test achievability of a given rate vector over $\mbb F_2$}
\end{center}
\end{table}
\end{exmp}

\section{Conclusion and Future Work}
This article defined the existential and enumerative variants of the constrained linear representability problem for polymatroids, and showed that special instances of these problems include the construction of achievability proofs in network coding and secret sharing.  An algorithm built from group theoretic techniques for combinatorial generation was developed to solve this problem, and an implementation of this algorithm in the GAP package \tsf{ITAP} accompanies the article.  Several experiments with the developed enumerative method demonstrated its utility as well as improvement in runtime over competing methods for solving CLRP$_q$-EX in some problems.  As network coding and secret sharing constructions in general necessitate nonlinear codes, a key future extension of the work will focus on finding nonlinear achievability constructions when linear ones fail.  In this vein, the group theoretic method of combinatorial generation employed here, Leiterspiel, can also be adapted to efficiently generate such nonlinear dependence structures, as demonstrated in \cite{Liu_JSTSP_EntGeoInfGeo_12,Liu_PhDdissertation}.

\section*{Acknowledgments}
\noindent This work was supported in part by the National Science Foundation under the awards CCF-1016588 and CCF-1421828.
\appendices
\section{Transformation of a HMSNC instance into a NCDAMG instance}\label{app:transform}
\begin{figure}[h]
\begin{center}
\includegraphics[width=0.8\textwidth]{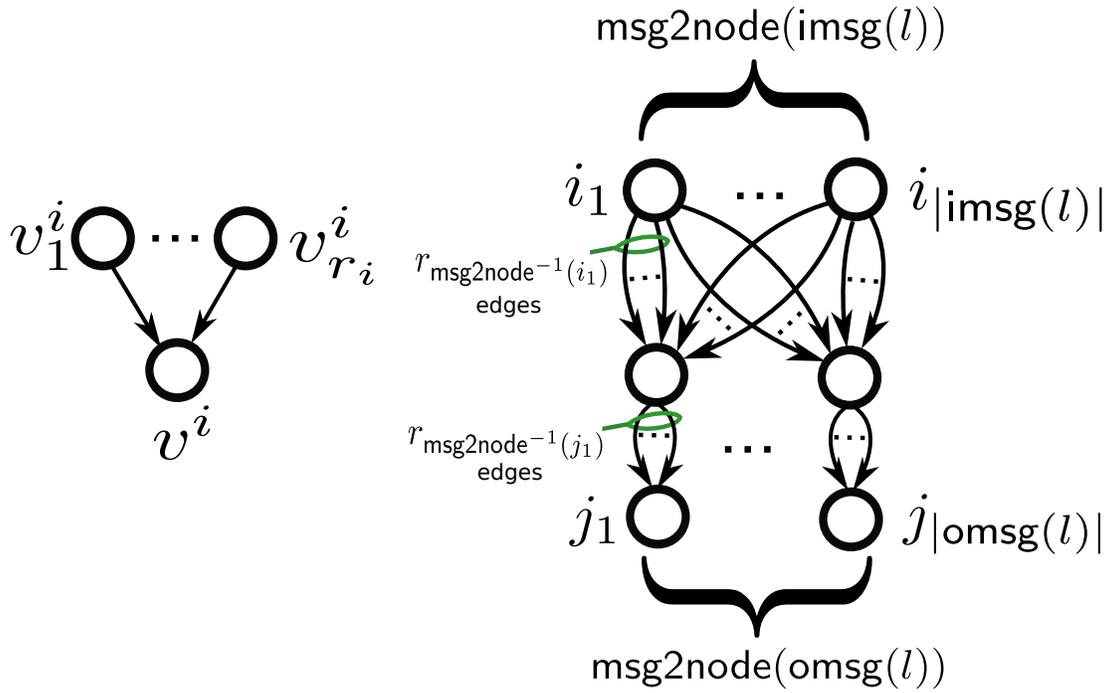}
\end{center}
\caption{(left) Gadget used by algorithm \ref{transformhmsnc} for source messages and, (right) gadget used for intermmediate constraints $l\in \mc L_2$}
\label{alg1g1}
\end{figure}

We specify a NCDAMG instance $A'$ as per the terminology in \cite{subrapathgain10}, by a tuple $(\mathscr V',\mathscr  E',\mathscr  S',\mathscr  T',g)$ where $(\mathscr V',\mathscr E')$ is a directed acyclic multigraph, $\mathscr S',\mathscr T'\subseteq V'$ are sets of source and sink nodes respectively, and $g$ is the demand function assigning exactly one member of $\mathscr S $ to each $t\in \mathscr T$. Each edge $e\in \mathscr E'$ is a triplet $(v_1,v_2,c)$ where $v_1,v_2\in \mathscr V'$ and $c$ is the edge label or color. Algorithm \ref{transformhmsnc} accepts a HMSNC instance in form of the associated constraints $\mc L_i,i\in[3]$, number of source $k$, number of random variables $N$ and a rate vector $(r_1,\hdots,r_N)$. At the begining, all members of $A'$ are empty.  Algorithm \ref{transformhmsnc} populates various member of $A'$ and also maintains a function $\tsf{msg2node}:\mathscr X\rightarrow \mathscr V'$ where $\mathscr X\subseteq [N]$. For message labels $i$ not in $\mathscr X$ at any point, we say that $\tsf{msg2node}(i)$ is not defined.  For each constraint $l$ in $\mc L_2\cup \mc L_3$, we use functions $\tsf{imsg}(l)$ and $\tsf{omsg}(l)$ to refer to the input and output message labels involved in constraint $l$. At the begining $\tsf{msg2node}(i)$ is undefined for each $i\in [N]$. First for each source message $i\in [k]$, $r_i$ nodes are added to sets $\mathscr V'$ and $\mathscr S'$ (lines \ref{alg1:l6}-\ref{alg1:l9}, fig. \ref{alg1g1}). These are the source nodes of NCDAMG instance $A'$.  Second, it goes through constraints $\mc L_2$, considering at each iteration of the while loop at lines \ref{alg1:l10}-\ref{alg1:l22}, a constraint $l$ s.t. $\tsf{msg2node}(i)$ is defined for each $i\in\tsf{imsg(l)}$.  The new nodes and edges, added to $\mathscr V'$ and $\mathscr E'$ using procedure \ref{proc:conl2}, are summarized as a gadget in figure \ref{alg1g1}. Finally, for each decoding constraint $l\in\mc L_3$, if $\vert \tsf{omsg}(l)\vert=1$ a single deoder node is added whereas if $\vert \tsf{omsg}(l)\vert>1$ multiple decoder nodes are added to $\mc V'$, in while loop at lines \ref{alg1:dec1}-\ref{alg1:dec2} using procedure \ref{proc:conl3}. The demand function $g$  for these decoder nodes is also set during the same procedure. Throughout algorithm \ref{transformhmsnc}, the number of parallel edges added between nodes depends on the rate vector. Thus, given a HMSNC instance $A$ and a rate vector $(r_1,\hdots,r_N)$, algorithm \ref{transformhmsnc} constructs a NCDAMG instance $A'$ such that the following holds:
\begin{lem} 
Rate vector $(r_1,\hdots,r_N)$ is achievable in $A$ with vector linear network codes if $A'$ is scalar linear solvable. 
\end{lem}
\textbf{Proof}: If $A'$ is scalar linear solvable, there exists a representable matroid $M$ associated with the scalar linear solution. In the matrix representation of $M$, each column is associated with an edge in $A'$. We can create a $N$-subspace arrangement $\{V_1,\hdots,V_N\}$ from this matroid s.t. $V_i,i\in [N]$ is the span of columns associated with edge incoming to $\tsf{msg2node}(i)$. $\blacksquare$\\
\begin{figure}[h]
\begin{center}
\includegraphics[width=0.8\textwidth]{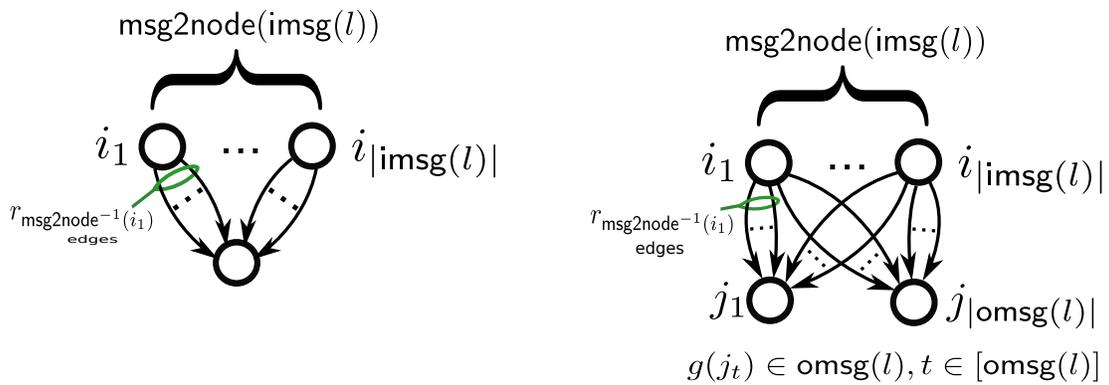}
\end{center}
\caption{Gadgets used by algorithm \ref{transformhmsnc} for decoder constraints. (left) the case $\vert\tsf{omsg}(l)\vert=1$ and (right) the case $\vert\tsf{omsg}(l)\vert>1$}
\label{alg1g2}
\end{figure}

 \begin{algorithm}[h!]
\caption{Algorithm to transform a HMSNC instance to a NCDAMG instance}
\label{transformhmsnc}
\DontPrintSemicolon 
\KwIn{Sets $\mc L_1,\mc L_2$ and $\mc L_3$ associated with HMSNC instance $A$, no. of sources k, no. of variables N, rate vector $\mb r=(r_1,\hdots,r_N)$}
\KwOut{A NCDAMG instance $A'=(\mathscr V',\mathscr  E',\mathscr  S',\mathscr  T',g)$}
$\mathscr V'\leftarrow \emptyset$, $\mathscr E'\leftarrow \emptyset$, $g\leftarrow$ empty function, $\tsf{msg2node}\leftarrow$ empty function, $\mc L\leftarrow \emptyset$\;
\ForEach{$i\in [k]$\label{alg1:l6}}
{
	$\mathscr V'\leftarrow \mathscr V'\cup\{v^i_1,\hdots,v^i_{r_i}\}$, $\mathscr S'\leftarrow \mathscr S'\cup\{v^i_1\hdots,v^i_{r_i}\}$, $\mathscr V'\leftarrow \mathscr V'\cup v^i$,	$\tsf{msg2node}(i)\leftarrow v^i$\;
}\label{alg1:l9}
\While{$\mc L_2\not\subseteq \mc L$\label{alg1:l10}}
{
	$l\leftarrow$ constraint in $\mc L_2$ not in $\mc L$ with $\tsf{msg2node}(i)$ defined for each $i\in \tsf{imsg}(l)$\;
	$(\tsf{msg2node},\mathscr V',\mathscr E')\leftarrow convertL2(l, \tsf{msg2node},\mathscr V',\mathscr E',\mb r)$\;
	$\mc L\leftarrow \mc L\cup\{l\}$
}\label{alg1:l22}
\While{$\mc L_3\not \subseteq \mc L$\label{alg1:dec1}}
{
	$l\leftarrow$ constraint in $\mc L_3$ not in $\mc L$\;
	$(\tsf{msg2node},\mathscr V',\mathscr E',\mathscr  T',g)\leftarrow convertL3(l, \tsf{msg2node},\mathscr V',\mathscr E',\mathscr  T',g,,\mb r)$\;
	$\mc L\leftarrow \mc L\cup\{l\}$
}\label{alg1:dec2} 
\Return{ $(\mathscr V',\mathscr  E',\mathscr  S',\mathscr  T',g)$}
\end{algorithm}

\begin{procedure} [h]
\DontPrintSemicolon
\caption{convertL2($l, \tsf{msg2node},\mathscr V',\mathscr E',,\mb r$) }\label{proc:conl2}
	\ForEach{$o\in \tsf{omsg}(l)$}
	{
		$\mathscr V'\leftarrow \mathscr V'\cup \{v^o_1\}$, $\mathscr V'\leftarrow \mathscr V'\cup \{v^o_2\}$, $\tsf{msg2node}(o)\leftarrow v^o_2$\;
				\ForEach{$i \in \tsf{imsg}(l)$}
		{
				$\mathscr E'\leftarrow \mathscr E'\cup\{(\tsf{msg2node}(i),v^o_1,c_1),\hdots, (\tsf{msg2node}(i), v^o_1,c_{r_i})\}$\;
		}
		$\mathscr E'\leftarrow \mathscr E'\cup\{(v^o_1,v^o_2,c_1),\hdots,(v^o_1,v^o_2,c_{r_o})\}$
	}
	\Return{$\tsf{msg2node},\mathscr V',\mathscr E'$}
\end{procedure}

\begin{procedure} [h]
\DontPrintSemicolon
\caption{convertL3($l, \tsf{msg2node},\mathscr V',\mathscr E', \mathscr T, g,\mb r$) }\label{proc:conl3}
	\If{$\vert \tsf{omsg}(l)\vert$ is 1}
	{
	$\mathscr V'\leftarrow \mathscr V'\cup \{v^l\}$\;
		\ForEach{$i \in \tsf{imsg}(l)$}
		{
				$\mathscr E'\leftarrow \mathscr E'\cup\{(\tsf{msg2node}(i),v^l,c_1),\hdots, (\tsf{msg2node}(i), v^l,c_{r_i})\}$\;
$g(v^l)\leftarrow \tsf{omsg}(l)$
		}
	$\mc T'\leftarrow \mathscr T'\cup \{v^l\}$
	}
	\Else
	{
		\ForEach{$o\in \tsf{omsg}(l)$}
		{
			$\mathscr V'\leftarrow \mathscr V'\cup \{v^l_o\}$\;
			\ForEach{$i \in \tsf{imsg}(l)$}
			{
				$\mathscr E'\leftarrow \mathscr E'\cup\{(\tsf{msg2node}(i),v^l_o,c_1),\hdots, (\tsf{msg2node}(i), v^l_o,c_{r_i})\}$\;
				$g(v^l_o)\leftarrow \{o\}$
			}
			$\mathscr T'\leftarrow \mathscr T'\cup \{v^l_o\}$
		}
	}
	\Return{$\tsf{msg2node},\mathscr V',\mathscr E', \mathscr T, g$}
\end{procedure}

\clearpage

\bibliographystyle{plain}
\bibliography{IEEEabrv,Allerton2015,myPubsWLinks}
\end{document}